\documentclass{ws-ijmpc}
\usepackage[super,compress]{cite}
\usepackage[breaklinks]{hyperref}
\hypersetup{colorlinks,urlcolor=black,citecolor=black,linkcolor=black,filecolor=black}
\usepackage{breakurl}
\usepackage{amsmath,amsfonts}
\usepackage{tikz}

\graphicspath{./all.the.plots/}

\def\db{\delta_b}
\def\dc{\delta_c}
\def\lp{{\ell}_{\rm Pl}}

\def\bib{B\kern-.05em{I}\kern-.025em{B}\kern-.08em}
\def\btex{B\kern-.05em{I}\kern-.025em{B}\kern-.08em\TeX}

\begin{document}

\markboth{A. Yonika, A. Heryudono, \& G. Khanna}
{Space-Time Collocation Method: Loop Quantum Hamiltonian Constraints}

\catchline{}{}{}{}{}

\title{Space-Time Collocation Method:
\\ Loop Quantum Hamiltonian Constraints
}

\author{A. Yonika}

\address{Department of Physics \& Center for Scientific Computing and Visualization Research, University of Massachusetts Dartmouth, 285 Old Westport Road\\
North Dartmouth, MA 02747,
USA\\
ayonika@umassd.edu}

\author{A. Heryudono}

\address{Department of Mathematics \& Center for Scientific Computing and Visualization Research, University of Massachusetts Dartmouth, 285 Old Westport Road\\
North Dartmouth, MA 02747,
USA\\
aheryudono@umassd.edu}

\author{G. Khanna}

\address{Department of Physics \& Center for Scientific Computing and Visualization Research, University of Massachusetts Dartmouth, 285 Old Westport Road\\
North Dartmouth, MA 02747,
USA\\
gkhanna@umassd.edu}
\maketitle

\begin{history}
\received{14 February 2020}
\revised{Day Month Year}
\end{history}

\begin{abstract}
A space-time collocation method (STCM) using asymptotically-constant basis functions is proposed and applied to the quantum Hamiltonian constraint for a loop-quantized treatment of the Schwarzschild interior. Canonically, these descriptions take the form of a partial-difference equation (PDE). The space-time collocation approach presents a computationally efficient, convergent, and easily parallelizable method for solving this class of equations, which is the main novelty of this study. Results of the numerical simulations will demonstrate the benefit from a parallel computing approach; and show general flexibility of the framework to handle arbitrarily-sized domains. Computed solutions will be compared, when applicable, to a solution computed in the conventional method via iteratively stepping through a predefined grid of discrete values, computing the solution via a recursive relationship.

\end{abstract}

\ccode{PACS Nos.: 02.70.-c, 04.60.-m}

\section{Introduction}
In loop quantum gravity the continuous structure of classical spacetime is replaced with a discrete structure arising from quantum geometric operators~[\refcite{thiemann2001introduction,Rovelli_1998,ashtekar2017overview}]. This trend continues in the symmetry-reduced models of loop quantum cosmology (LQC). The consequential discrete structure of spacetime is only applicable when the curvature approaches Planck scale, otherwise, agreement with general relativity is expected and observed. A large body of work has been dedicated towards the problem of singularity resolution in the symmetry-reduced cosmological models. There it is found that the big bang is replaced by a big bounce~[\refcite{Ashtekar_2011,Bojowald_2005,Ashtekar_20061,Ashtekar_20062}], and the loop-quantized Hamiltonian constraints are replaced by {\em partial difference equations}. These discrete equations describe the quantum gravity evolution of spacetime in various models, such as cosmological or Schwarzschild~[\refcite{Brizuela_2012,Singh_2012}]. Typically, numerical simulations of the relevant Hamiltonian constraint equations are obtained in an iterative stepping manner. Initial conditions of symmetric Gaussian wave packets are imposed, representing a quantum wave function i.e. a probabilistic distribution of the state of the system. These initial conditions are then evolved through a predefined discrete spacetime grid, in a process reminiscent of iterative finite-difference methods for the numerical solution of differential equations.

Despite preserving the exact nature of the Hamiltonian constraints, the detriments of finding a solution using this iterative-stepping process are well-documented~ [\refcite{Singh_2018,Saini_2019,Yonika_2018}]. Large domains quickly become infeasible to compute, due to an increasing demand for higher levels of computational precision~[\refcite{Yonika_2018}]. Any challenges regarding the required computational resources become compounded by the fact that the recursive nature of these equations make a parallel implementation very difficult. Lately, attention has been paid towards finding accurate-yet-efficient means to compute solutions on larger, more physically-viable, domains. Success in applying a collocation-inspired method over a single variable was recently demonstrated, showing benefit over the iterative stepping method in both permissible precision and parallelizability~[\refcite{Yonika_2019}].

The idea of space-time collocation schemes on a tensor product grid for numerically solving partial differential equations (PDEs) is not new. One can trace back the idea to the 70s with the work of Douglas and Dupont [\refcite{douglasdupont74}] (see chapter II, section 11, entitled collocation is space and time). Space-time formulation treats time as another space variable, and the discretized PDEs or initial boundary value problems (IBVPs) are solved entirely as boundary value problems in the space-time domain. Since then, researchers have modified the method with several variants, particularly in the choices of bases in space and time directions. 

The tensor-product space-time collocation scheme is considered due to the following purposes:
\begin{enumerate}
\item Unlike the method of lines, the space-time BVP solver avoids the problem in choosing the right ODE solver that depends on PDE types (e.g., parabolic, hyperbolic, or mixed types.). This benefits for problems with high-degree simultaneous space-time intermittency. For such problems, the goal is to avoid having unbalanced accuracies in space and time solutions.
\item The method offers flexibility in choosing the bases for spatial and temporal approximations. It also leads to freedom in using a different layout of the tensor-product grid to mimic to the solution profiles.
\item Techniques for implementing boundary conditions can be done in the same way as standard BVP for time-independent problems.
\item The space-time domain can also be divided into multiple space-time ``slabs.'' The top of one slab provides an initial condition at the bottom of the next slab so that the process resembles typical time-marching. 
\end{enumerate}
However, the method is not without drawbacks. The includes:
\begin{enumerate}
\item The method is implicit and involves solving big linear or non-linear systems. Depending on the problems and the treatment of boundary conditions, the system matrix may not have special properties that can be easily exploited to speed up computing the solutions.  
\item Unlike the method of lines, a theoretical understanding of the method is less explored. First, finding a preconditioner to solve the system iteratively is non-trivial. Second, the stability of the method is not widely understood. Third, the solvability of the problem may not be resulting in the right solution.
\item Using the tensor-product grid in non-regular static or moving domains is non-trivial. In this case, a space-time method based on finite element or radial basis functions can be a viable choice.
\end{enumerate}
For the particular problem provided in this paper, we chose space-time formulation due to its advantages mentioned above. In order to deal with the ill-conditioning issue, we use the least-squares approach, which results in more rows than columns in the system matrix. In this paper, we focus on the practical aspect of the space-time method for our particular problem. Theoretical work will be left for future studies.

This paper is organized as follows: Sec.~\ref{ST-Mat} discusses the generalized process behind formulating a problem in the context of a STCM. In Sec.~\ref{CS-M} we introduce the Corichi and Singh (CS) model for the loop-quantized Schwarzschild interior, which will be the focus of our implementation of the STCM. Applying the STCM towards the CS model will be demonstrated in Sec.~\ref{ST-CS}, along with a discussion of pertinent modifications shown to be beneficial towards computing physically viable solutions in Sec.~\ref{ST-CS-B} and Sec.~\ref{ST-CS-LS}. Our results for the CS model using the STCM will be presented in Sec.~\ref{Results}. These results will be compared with solutions computed via iterative-stepping in Sec.~\ref{R-E}; and we will remark on characteristics of the STCM solutions in Sec.~\ref{R-P} \& Sec.~\ref{R-A}.




\section{Tensor product grid space-time collocation scheme}\label{ST-Mat}
\begin{figure}[htp]
\begin{minipage}{\textwidth}
\begin{center}
\begin{tikzpicture}
\draw (1,0) -- (4,0) -- (4,2) -- (1,2) -- cycle;
\draw (2.5,0) node[label=below:$x$](b){};
\draw (1,1) node[label=left:$t$](b){};
\draw (6,0) -- (9,0) -- (9,2) -- (6,2) -- cycle;
\draw[step=0.5cm,black] (6,0) grid (9,2);
\draw (6,2) node[circle, fill, inner sep=1pt, label=above left:$x_{11}$](b){};
\draw (6,1.5) node[circle, fill, inner sep=1pt, label=left:$x_{21}$](b){};
\draw (6,1) node[circle, fill, inner sep=1pt, label=left:$x_{31}$](b){};
\draw (6,0.5) node[circle, fill, inner sep=1pt, label=left:$\vdots$](b){};
\draw (6,0) node[circle, fill, inner sep=1pt, label=below left:$x_{m1}$](b){};

\draw (6.5,2) node[circle, fill, inner sep=1pt, label=above:$x_{12}$](b){};
\draw (7,2) node[circle, fill, inner sep=1pt, label=above:$ $](b){};
\draw (7.5,2) node[circle, fill, inner sep=1pt, label=above:$\dots$](b){};
\draw (8,2) node[circle, fill, inner sep=1pt, label=above:$\dots$](b){};
\draw (8.5,2) node[circle, fill, inner sep=1pt, label=above:$ $](b){};
\draw (9,2) node[circle, fill, inner sep=1pt, label=above right:$x_{1n}$](b){};

\draw (9,1.5) node[circle, fill, inner sep=1pt, label=right:$x_{2n}$](b){};
\draw (9,1) node[circle, fill, inner sep=1pt, label=right:$x_{3n}$](b){};
\draw (9,0.5) node[circle, fill, inner sep=1pt, label=right:$\vdots$](b){};
\draw (9,0) node[circle, fill, inner sep=1pt, label=below right:$x_{mn}$](b){};

\draw (6.5,0) node[circle, fill, inner sep=1pt, label=below:$x_{m2}$](b){};
\draw (7,0) node[circle, fill, inner sep=1pt, label=below:$ $](b){};
\draw (7.5,0) node[circle, fill, inner sep=1pt, label=below:$\dots$](b){};
\draw (8,0) node[circle, fill, inner sep=1pt, label=below:$\dots $](b){};
\draw (8.5,0) node[circle, fill, inner sep=1pt, label=below:$ $](b){};

\draw (11,0) -- (14,0) -- (14,2) -- (11,2) -- cycle;
\draw[step=0.5cm,black] (11,0) grid (14,2);
\draw (11,2) node[circle, fill, inner sep=1pt, label=above left:$t_{11}$](b){};
\draw (11,1.5) node[circle, fill, inner sep=1pt, label=left:$t_{21}$](b){};
\draw (11,1) node[circle, fill, inner sep=1pt, label=left:$t_{31}$](b){};
\draw (11,0.5) node[circle, fill, inner sep=1pt, label=left:$\vdots$](b){};
\draw (11,0) node[circle, fill, inner sep=1pt, label=below left:$t_{m1}$](b){};

\draw (11.5,2) node[circle, fill, inner sep=1pt, label=above:$t_{12}$](b){};
\draw (12,2) node[circle, fill, inner sep=1pt, label=above:$ $](b){};
\draw (12.5,2) node[circle, fill, inner sep=1pt, label=above:$\dots$](b){};
\draw (13,2) node[circle, fill, inner sep=1pt, label=above:$\dots$](b){};
\draw (13.5,2) node[circle, fill, inner sep=1pt, label=above:$ $](b){};
\draw (14,2) node[circle, fill, inner sep=1pt, label=above right:$t_{1n}$](b){};

\draw (14,1.5) node[circle, fill, inner sep=1pt, label=right:$t_{2n}$](b){};
\draw (14,1) node[circle, fill, inner sep=1pt, label=right:$t_{3n}$](b){};
\draw (14,0.5) node[circle, fill, inner sep=1pt, label=right:$\vdots$](b){};
\draw (14,0) node[circle, fill, inner sep=1pt, label=below right:$t_{mn}$](b){};

\draw (11.5,0) node[circle, fill, inner sep=1pt, label=below:$t_{m2}$](b){};
\draw (12,0) node[circle, fill, inner sep=1pt, label=below:$ $](b){};
\draw (12.5,0) node[circle, fill, inner sep=1pt, label=below:$\dots$](b){};
\draw (13,0) node[circle, fill, inner sep=1pt, label=below:$\dots $](b){};
\draw (13.5,0) node[circle, fill, inner sep=1pt, label=below:$ $](b){};
\end{tikzpicture}
\end{center}
\end{minipage}
\caption{An example of a tensor product grid layout of a 1D+t bounded rectangular domain. Coordinate values of $x$ and $t$ can be stored as matrices.}
\label{fig:stgrid}
\end{figure}
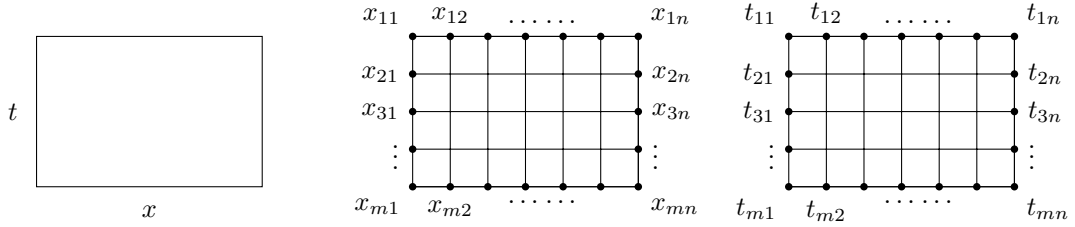

In a space-time collocation scheme, time is treated as another space variable. Our goal is to numerically solve the discretized PDE entirely as a space-time BVP. We start by discretizing the domain. As an example, Figure \ref{fig:stgrid} shows a tensor-product grid layout of a rectangular (1D+t) bounded domain. The spatial $x$ and temporal $t$ axes are divided into $n$ and $m$ points, respectively, resulting in a rectangular grid. As a result, the discrete coordinate values of $x$ and $t$ can be stored as $m \times n$ matrices.

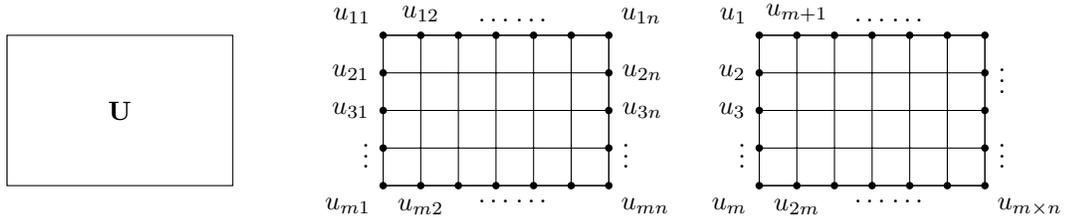
\begin{figure}[htp]
\begin{minipage}{\textwidth}
\begin{center}
\begin{tikzpicture}
\draw (1,0) -- (4,0) -- (4,2) -- (1,2) -- cycle;
\draw (2.5,1) node[label=center:$\mathbf{U}$](b){};
\draw (6,0) -- (9,0) -- (9,2) -- (6,2) -- cycle;
\draw[step=0.5cm,black] (6,0) grid (9,2);
\draw (6,2) node[circle, fill, inner sep=1pt, label=above left:$u_{11}$](b){};
\draw (6,1.5) node[circle, fill, inner sep=1pt, label=left:$u_{21}$](b){};
\draw (6,1) node[circle, fill, inner sep=1pt, label=left:$u_{31}$](b){};
\draw (6,0.5) node[circle, fill, inner sep=1pt, label=left:$\vdots$](b){};
\draw (6,0) node[circle, fill, inner sep=1pt, label=below left:$u_{m1}$](b){};

\draw (6.5,2) node[circle, fill, inner sep=1pt, label=above:$u_{12}$](b){};
\draw (7,2) node[circle, fill, inner sep=1pt, label=above:$ $](b){};
\draw (7.5,2) node[circle, fill, inner sep=1pt, label=above:$\dots$](b){};
\draw (8,2) node[circle, fill, inner sep=1pt, label=above:$\dots$](b){};
\draw (8.5,2) node[circle, fill, inner sep=1pt, label=above:$ $](b){};
\draw (9,2) node[circle, fill, inner sep=1pt, label=above right:$u_{1n}$](b){};

\draw (9,1.5) node[circle, fill, inner sep=1pt, label=right:$u_{2n}$](b){};
\draw (9,1) node[circle, fill, inner sep=1pt, label=right:$u_{3n}$](b){};
\draw (9,0.5) node[circle, fill, inner sep=1pt, label=right:$\vdots$](b){};
\draw (9,0) node[circle, fill, inner sep=1pt, label=below right:$u_{mn}$](b){};

\draw (6.5,0) node[circle, fill, inner sep=1pt, label=below:$u_{m2}$](b){};
\draw (7,0) node[circle, fill, inner sep=1pt, label=below:$ $](b){};
\draw (7.5,0) node[circle, fill, inner sep=1pt, label=below:$\dots$](b){};
\draw (8,0) node[circle, fill, inner sep=1pt, label=below:$\dots $](b){};
\draw (8.5,0) node[circle, fill, inner sep=1pt, label=below:$ $](b){};

\draw (11,0) -- (14,0) -- (14,2) -- (11,2) -- cycle;
\draw[step=0.5cm,black] (11,0) grid (14,2);
\draw (11,2) node[circle, fill, inner sep=1pt, label=above left:$u_{1}$](b){};
\draw (11,1.5) node[circle, fill, inner sep=1pt, label=left:$u_{2}$](b){};
\draw (11,1) node[circle, fill, inner sep=1pt, label=left:$u_{3}$](b){};
\draw (11,0.5) node[circle, fill, inner sep=1pt, label=left:$\vdots$](b){};
\draw (11,0) node[circle, fill, inner sep=1pt, label=below left:$u_{m}$](b){};

\draw (11.5,2) node[circle, fill, inner sep=1pt, label=above:$u_{m+1}$](b){};
\draw (12,2) node[circle, fill, inner sep=1pt, label=above:$ $](b){};
\draw (12.5,2) node[circle, fill, inner sep=1pt, label=above:$\dots$](b){};
\draw (13,2) node[circle, fill, inner sep=1pt, label=above:$\dots$](b){};
\draw (13.5,2) node[circle, fill, inner sep=1pt, label=above:$ $](b){};
\draw (14,2) node[circle, fill, inner sep=1pt, label=above right:$$](b){};

\draw (14,1.5) node[circle, fill, inner sep=1pt, label=right:$\vdots$](b){};
\draw (14,1) node[circle, fill, inner sep=1pt, label=right:$$](b){};
\draw (14,0.5) node[circle, fill, inner sep=1pt, label=right:$\vdots$](b){};
\draw (14,0) node[circle, fill, inner sep=1pt, label=below right:$u_{m\times n}$](b){};

\draw (11.5,0) node[circle, fill, inner sep=1pt, label=below:$u_{2m}$](b){};
\draw (12,0) node[circle, fill, inner sep=1pt, label=below:$ $](b){};
\draw (12.5,0) node[circle, fill, inner sep=1pt, label=below:$\dots$](b){};
\draw (13,0) node[circle, fill, inner sep=1pt, label=below:$\dots $](b){};
\draw (13.5,0) node[circle, fill, inner sep=1pt, label=below:$ $](b){};
\end{tikzpicture}
\end{center}
\end{minipage}
\caption{An example of solution matrix $\mathbf{U}$ in matrix and linear indices.}
\label{fig:stsol}
\end{figure}

The solution $u$, which is a function of $x$ and $t$, is also stored as an $m \times n$ matrix. However, as shown in the rightmost figure of Figure \ref{fig:stsol}, it is also common to vectorize the matrix $\mathbf{U}$ by stacking its columns and order it in lexicographic (with linear indices) way and avoid using two subindices, i.e., $u_{ij}$. Hence, $\underline{u}=\mathtt{vec}(\mathbf{U})=[u_1,\dots,u_{m\times n}]^T$. The vectorized version of matrices $\mathbf{X}$ and $\mathbf{T}$ can be obtained in the same way if needed.   

Computing the derivatives of $u$ with respect to $x$ or $t$ only involves a product of $\mathbf{U}$ with suitable differentiation matrices. Given an $m \times m$ differentiation matrix $\mathbf{D}_t$ that represents the discrete derivative operator with respect to $t$, the $m \times n$ matrix $\mathbf{U}_t$ that stores the discrete values of $u_t$ can be computed as  

\begin{align}
    \mathbf{U}_t = \mathbf{D}_t \mathbf{U}.
\end{align}

In short, to compute $\mathbf{U}_t$, we multiply the matrix $\mathbf{D}_t$ to every column of $\mathbf{U}$. On the other hand, the $m \times n$ matrix $\mathbf{U}_x$ that stores the discrete values of $u_x$ can be computed as   

\begin{align}
    \mathbf{U}_x = \mathbf{U} \mathbf{D}_x^T,
\end{align}
where $\mathbf{D}_x$ is an $n \times n$ differentiation matrix that represents the spatial derivative operator. Depending on the methods, the matrices $\mathbf{D}_t$ and $\mathbf{D}_x$ can be generated using finite-difference, pseudospectral, radial basis function, or other methods. Higher derivatives $\mathbf{U}_{xx},\mathbf{U}_{tt},\dots$ can be computed in the same way by using the appropriate high-derivative matrix operators $\mathbf{D}_{xx},\mathbf{D}_{tt},\dots$. 

A connection of $\underline{u}_t = \mathtt{vec}(\mathbf{U}_t)$ and $\underline{u}_x = \mathtt{vec}(\mathbf{U}_x)$ with $\underline{u} = \mathtt{vec}(\mathbf{U})=[u_1,\dots,u_{m\times n}]^T$ can be described via kronecker tensor product operator. As an example, the vector $\underline{u_t}$ is
\begin{align}
   \underline{u_t} = (\mathbf{I}_n \otimes \mathbf{D}_t) \underline{u},
\end{align}
where $\mathbf{I}_n$ is an $n \times n$ identity matrix and $\mathbf{I}_n \otimes \mathbf{D}_t$ is a square block diagonal matrix of size $(m \times n) \times (m \times n)$. Additionally, the vector $\underline{u_x}$ can be computed as
\begin{align}
   \underline{u_x} = (\mathbf{D}_x \otimes \mathbf{I}_m) \underline{u},
\end{align}
where $\mathbf{I}_m$ is an $m \times m$ identity matrix.

When solving a time-dependent PDE as a space-time BVP, our goal is to compute the unknown $\underline{u}$. In the space-time BVP, the initial condition is treated as a Dirichlet type boundary condition. We will see later that implementing boundary conditions is done by modifying the appropriate rows of the system matrix. As a simple example, suppose we want to solve a linear BVP problem given by
\begin{align}
\displaystyle \frac{\partial u}{\partial x} + \frac{\partial u}{\partial t} &= f(x,t), \quad (x,t) \in \Omega \cup \partial \Omega_\textnormal{north} \cup \Omega_\textnormal{west},\\
u(x,t) &= g(x,t), \quad (x,t) \in \partial \Omega_\textnormal{south} \cup \partial \Omega_\textnormal{east},
\end{align}
where $\overline{\Omega} = \{a \leq x \leq b, t_i \leq t \leq t_f\}$. In order to simply visualize the discretization of the BVP without using too many grid points, let us discretize $x$ into $n = 4$ and sample $t$ with $m=3$ points.

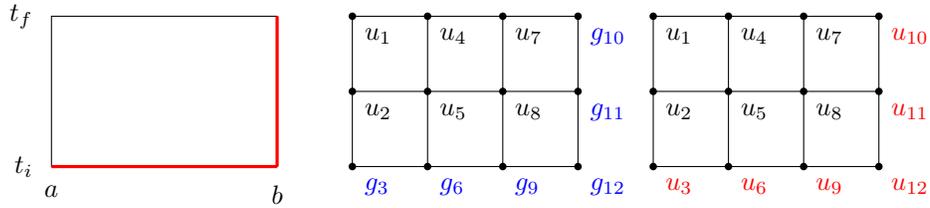
\begin{figure}[htp]
\begin{minipage}{\textwidth}
\begin{center}
\begin{tikzpicture}
\draw (-6,0) -- (-3,0) -- (-3,2) -- (-6,2) -- cycle;
\draw[red,very thick] (-6,0) -- (-3,0);
\draw[red,very thick] (-3,0) -- (-3,2);
\draw (-6,0) node[label=below:$a$](b){};
\draw (-6,2) node[label=left:$t_f$](b){};
\draw (-3,0) node[label=below:$b$](b){};
\draw (-6,0) node[label=left:$t_i$](b){};

\draw[step=1cm,black] (-2,0) grid (1,2);
\draw (-2,0) node[circle, fill, inner sep=1pt, label=below right:$\textcolor{blue}{g_{3}}$](b){};
\draw (-1,0) node[circle, fill, inner sep=1pt, label=below right:$\textcolor{blue}{g_{6}}$](b){};
\draw (0,0) node[circle, fill, inner sep=1pt, label=below right:$\textcolor{blue}{g_{9}}$](b){};
\draw (1,0) node[circle, fill, inner sep=1pt, label=below right:$\textcolor{blue}{g_{12}}$](b){};
\draw (-2,1) node[circle, fill, inner sep=1pt, label=below right:$u_{2}$](b){};
\draw (-1,1) node[circle, fill, inner sep=1pt, label=below right:$u_{5}$](b){};
\draw (0,1) node[circle, fill, inner sep=1pt, label=below right:$u_{8}$](b){};
\draw (1,1) node[circle, fill, inner sep=1pt, label=below right:$\textcolor{blue}{g_{11}}$](b){};
\draw (-2,2) node[circle, fill, inner sep=1pt, label=below right:$u_{1}$](b){};
\draw (-1,2) node[circle, fill, inner sep=1pt, label=below right:$u_{4}$](b){};
\draw (0,2) node[circle, fill, inner sep=1pt, label=below right:$u_{7}$](b){};
\draw (1,2) node[circle, fill, inner sep=1pt, label=below right:$\textcolor{blue}{g_{10}}$](b){};

\draw[step=1cm,black] (2,0) grid (5,2);
\draw (2,0) node[circle, fill, inner sep=1pt, label=below right:$\textcolor{red}{u_3}$](b){};
\draw (3,0) node[circle, fill, inner sep=1pt, label=below right:$\textcolor{red}{u_6}$](b){};
\draw (4,0) node[circle, fill, inner sep=1pt, label=below right:$\textcolor{red}{u_9}$](b){};
\draw (5,0) node[circle, fill, inner sep=1pt, label=below right:$\textcolor{red}{u_{12}}$](b){};
\draw (2,1) node[circle, fill, inner sep=1pt, label=below right:$u_2$](b){};
\draw (3,1) node[circle, fill, inner sep=1pt, label=below right:$u_5$](b){};
\draw (4,1) node[circle, fill, inner sep=1pt, label=below right:$u_8$](b){};
\draw (5,1) node[circle, fill, inner sep=1pt, label=below right:$\textcolor{red}{u_{11}}$](b){};
\draw (2,2) node[circle, fill, inner sep=1pt, label=below right:$u_1$](b){};
\draw (3,2) node[circle, fill, inner sep=1pt, label=below right:$u_4$](b){};
\draw (4,2) node[circle, fill, inner sep=1pt, label=below right:$u_7$](b){};
\draw (5,2) node[circle, fill, inner sep=1pt, label=below right:$\textcolor{red}{u_{10}}$](b){};
\end{tikzpicture}
\end{center}
\end{minipage}
\caption{The south and east boundaries of the domain are the locations where the boundary conditions are enforced (shown in red). The initial condition at the south boundary is treated as Dirichlet boundary condition.}
\label{fig:stsoln4m3}
\end{figure}

Figure \ref{fig:stsoln4m3} shows that the south and east boundaries (in red color) of the domain are the locations where boundary conditions are enforced. At those points, the values of $u$ are known. The discrete representation of $\mathcal{L}u=\frac{\partial u}{\partial x} + \frac{\partial u}{\partial t}$ all the way to the boundaries becomes
\begin{align}
\mathbf{\mathbf{L}} \underline{u}=(\mathbf{D}_x \otimes \mathbf{I}_m + \mathbf{I}_n \otimes \mathbf{D}_t) \underline{u}  = (\mathbf{D}_x \otimes \mathbf{I}_3 + \mathbf{I}_4 \otimes \mathbf{D}_t) \underline{u},
\end{align}
where $\mathbf{D}_x$ and $\mathbf{D}_t$ are $4 \times 4$ and $3 \times 3$ square differentiation matrices respectively. The sparsity distribution of the $\mathbf{L}$ operator can be shown as

\begin{align}
\label{eq:spmatdist}
\mathbf{\mathbf{L}} \underline{u} = \left[
\begin{array}{c|c|c|c}
\begin{matrix} 
\bullet & \bullet &  \bullet \\
\bullet & \bullet &  \bullet \\
\textcolor{red}{\bullet} & \textcolor{red}{\bullet} &  \textcolor{red}{\bullet} \\
\end{matrix}
 &
\begin{matrix} 
\bullet &  &   \\
 & \bullet &   \\
 &   &  \textcolor{red}{\bullet} \\
\end{matrix} 
  & 
\begin{matrix} 
\bullet &  &   \\
 & \bullet &   \\
 &   &  \textcolor{red}{\bullet} \\
\end{matrix}   
  & \\ 
\hline
\begin{matrix} 
\bullet &  &   \\
 & \bullet &   \\
 &   &\textcolor{red}{\bullet} \\
\end{matrix} 
&
\begin{matrix} 
\bullet & \bullet &  \bullet \\
\bullet & \bullet &  \bullet \\
\textcolor{red}{\bullet} & \textcolor{red}{\bullet} &  \textcolor{red}{\bullet} \\
\end{matrix}
& 
\begin{matrix} 
\bullet &  &   \\
 & \bullet &   \\
 &   &  \textcolor{red}{\bullet} \\
\end{matrix} 
& \\
\hline
& 
\begin{matrix} 
\bullet &  &   \\
 & \bullet &   \\
 &   & \textcolor{red}{\bullet} \\
\end{matrix} 
&
\begin{matrix} 
\bullet & \bullet &  \bullet \\
\bullet & \bullet &  \bullet \\
\textcolor{red}{\bullet} & \textcolor{red}{\bullet} &  \textcolor{red}{\bullet} \\
\end{matrix}
& 
\begin{matrix} 
\bullet &  &   \\
 & \bullet &   \\
 &   &  \textcolor{red}{\bullet} \\
\end{matrix} 
\\
\hline
& 
\begin{matrix} 
\textcolor{red}{\bullet} &  &   \\
 & \textcolor{red}{\bullet} &   \\
 &   &  \textcolor{red}{\bullet} \\
\end{matrix} 
& 
\begin{matrix} 
\textcolor{red}{\bullet} &  &   \\
 & \textcolor{red}{\bullet} &   \\
 &   &  \textcolor{red}{\bullet} \\
\end{matrix} 
&
\begin{matrix} 
\textcolor{red}{\bullet} & \textcolor{red}{\bullet} &  \textcolor{red}{\bullet} \\
\textcolor{red}{\bullet} & \textcolor{red}{\bullet} &  \textcolor{red}{\bullet} \\
\textcolor{red}{\bullet} & \textcolor{red}{\bullet} &  \textcolor{red}{\bullet} \\
\end{matrix}
\\
\end{array}
\right]
\begin{bmatrix}
u_{1} \\ u_{2}  \\ \textcolor{red}{u_{3}} \\ 
u_{4} \\ u_{5}  \\ \textcolor{red}{u_{6}} \\ 
u_{7} \\ u_{8}  \\ \textcolor{red}{u_{9}} \\
 \textcolor{red}{u_{10}} \\ \textcolor{red}{u_{11}}  \\ \textcolor{red}{u_{12}} \\
\end{bmatrix}.
\end{align}

In order to enforce the Dirichlet boundary condition, we modify the rows containing red dots in \eqref{eq:spmatdist}. Hence, the system of linear equations along with the correct right-hand side vectors for solving $\underline{u}$ becomes

\begin{align}
\label{eq:spmatdistmodified}
\left[
\begin{array}{c|c|c|c}
\begin{matrix} 
\bullet & \bullet &  \bullet \\
\bullet & \bullet &  \bullet \\
          &            & \textcolor{red}{1} \\
\end{matrix}
 &
\begin{matrix} 
\bullet &  &   \\
 & \bullet &   \\
 &   &  \\
\end{matrix} 
  & 
\begin{matrix} 
\bullet &  &   \\
 & \bullet &   \\
 &   &  \\
\end{matrix}   
  & \\ 
\hline
\begin{matrix} 
\bullet &  &   \\
 & \bullet &   \\
 &   & \\
\end{matrix} 
&
\begin{matrix} 
\bullet & \bullet &  \bullet \\
\bullet & \bullet &  \bullet \\
 &   & \textcolor{red}{1} \\
\end{matrix}
& 
\begin{matrix} 
\bullet &  &   \\
 & \bullet &   \\
 &   &  \\
\end{matrix} 
& \\
\hline
& 
\begin{matrix} 
\bullet &  &   \\
 & \bullet &   \\
 &   & \\
\end{matrix} 
&
\begin{matrix} 
\bullet & \bullet &  \bullet \\
\bullet & \bullet &  \bullet \\
 &   & \textcolor{red}{1} \\
\end{matrix}
& 
\begin{matrix} 
\bullet &  &   \\
 & \bullet &   \\
 &   &  \\
\end{matrix} 
\\
\hline
& 
& 
&
\begin{matrix} 
\textcolor{red}{1} &  &  \\
 & \textcolor{red}{1} &   \\
 &   & \textcolor{red}{1} \\
\end{matrix}
\\
\end{array}
\right]
\begin{bmatrix}
u_{1} \\ u_{2}  \\ \textcolor{red}{u_{3}} \\ 
u_{4} \\ u_{5}  \\ \textcolor{red}{u_{6}} \\ 
u_{7} \\ u_{8}  \\ \textcolor{red}{u_{9}} \\
 \textcolor{red}{u_{10}} \\ \textcolor{red}{u_{11}}  \\ \textcolor{red}{u_{12}} \\
\end{bmatrix}
=
\begin{bmatrix}
f_{1} \\ f_{2}  \\  \textcolor{red}{g_{3}} \\ 
f_{4} \\ f_{5}  \\  \textcolor{red}{g_{6}} \\ 
f_{7} \\ f_{8}  \\  \textcolor{red}{g_{9}} \\
 \textcolor{red}{g_{10}} \\  \textcolor{red}{g_{11}}  \\  \textcolor{red}{g_{12}} \\
\end{bmatrix}.
\end{align}

The unknown vector $\underline{u}$ can then be solved using a suitable system of linear equations solver. When the system matrix in \eqref{eq:spmatdistmodified} is ill-conditioned and finding a pre-conditioner is non-trivial, one may also extend the scheme in a least-squares sense. The discrete least-squares collocation technique for solving PDEs can be traced back to the technical report by Houstis [\refcite{houstis1978collocation}] in late 70s. In this case, extra grid points are added to collocate the BVP. The least-square approach leads to a rectangular system with more rows than columns as opposed to a square system in \eqref{eq:spmatdistmodified}.  

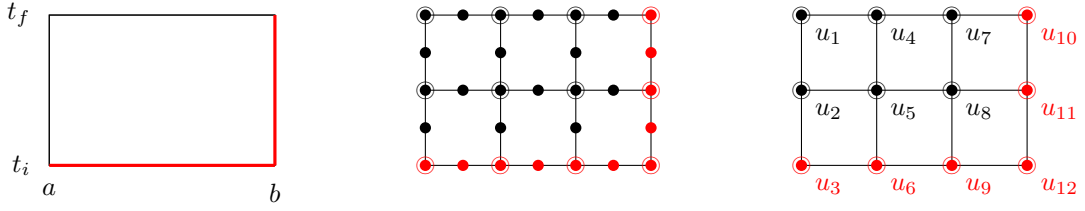
\begin{figure}[htp]
\begin{minipage}{\textwidth}
\begin{center}
\begin{tikzpicture}
\draw (1,0) -- (4,0) -- (4,2) -- (1,2) -- cycle;
\draw (1,0) -- (4,0) -- (4,2) -- (1,2) -- cycle;
\draw[red,very thick] (1,0) -- (4,0);
\draw[red,very thick] (4,0) -- (4,2);
\draw (1,0) node[label=below:$a$](b){};
\draw (1,2) node[label=left:$t_f$](b){};
\draw (4,0) node[label=below:$b$](b){};
\draw (1,0) node[label=left:$t_i$](b){};

\draw[step=1.0cm,black] (6,0) grid (9,2);
\draw (6,2) node[circle, fill, inner sep=1.5pt, label=center:$\odot$](b){};
\draw (6,1.5) node[circle, fill, inner sep=1.5pt, label=center:$$](b){};
\draw (6,1) node[circle, fill, inner sep=1.5pt, label=center:$\odot$](b){};
\draw (6,0.5) node[circle, fill, inner sep=1.5pt, label=center:$$](b){};
\draw (6,0) node[circle, red,fill, inner sep=1.5pt, label=center:$\color{red}{\odot}$](b){};

\draw (6.5,2) node[circle, fill, inner sep=1.5pt, label=center:$$](b){};
\draw (6.5,1) node[circle, fill, inner sep=1.5pt, label=center:$$](b){};
\draw (6.5,0) node[circle, red, fill, inner sep=1.5pt, label=center:$$](b){};

\draw (7,2) node[circle, fill, inner sep=1.5pt, label=center:$\odot$](b){};
\draw (7,1.5) node[circle, fill, inner sep=1.5pt, label=center:$$](b){};
\draw (7,1) node[circle, fill, inner sep=1.5pt, label=center:$\odot$](b){};
\draw (7,0.5) node[circle, fill, inner sep=1.5pt, label=center:$$](b){};
\draw (7,0) node[circle, red, fill, inner sep=1.5pt, label=center:$\color{red}{\odot}$](b){};

\draw (7.5,2) node[circle, fill, inner sep=1.5pt, label=center:$$](b){};
\draw (7.5,1) node[circle, fill, inner sep=1.5pt, label=center:$$](b){};
\draw (7.5,0) node[circle, red, fill, inner sep=1.5pt, label=center:$$](b){};

\draw (8,2) node[circle, fill, inner sep=1.5pt, label=center:$\odot$](b){};
\draw (8,1.5) node[circle, fill, inner sep=1.5pt, label=center:$$](b){};
\draw (8,1) node[circle, fill, inner sep=1.5pt, label=center:$\odot$](b){};
\draw (8,0.5) node[circle, fill, inner sep=1.5pt, label=center:$$](b){};
\draw (8,0) node[circle, red, fill, inner sep=1.5pt, label=center:$\color{red}{\odot}$](b){};

\draw (8.5,2) node[circle, fill, inner sep=1.5pt, label=center:$$](b){};
\draw (8.5,1) node[circle, fill, inner sep=1.5pt, label=center:$$](b){};
\draw (8.5,0) node[circle, red, fill, inner sep=1.5pt, label=center:$$](b){};

\draw (9,2) node[circle, red, fill, inner sep=1.5pt, label=center:$\color{red}{\odot}$](b){};
\draw (9,1.5) node[circle, red, fill, inner sep=1.5pt, label=center:$$](b){};
\draw (9,1) node[circle, red, fill, inner sep=1.5pt, label=center:$\color{red}{\odot}$](b){};
\draw (9,0.5) node[circle, red, fill, inner sep=1.5pt, label=center:$$](b){};
\draw (9,0) node[circle, red, fill, inner sep=1.5pt, label=center:$\color{red}{\odot}$](b){};

\draw[step=1cm,black] (11,0) grid (14,2);
\draw (11,0) node[circle, red, fill, inner sep=1.5pt, label=below right:$\textcolor{red}{u_3}$,label=center:$\color{red}{\odot}$](b){};
\draw (12,0) node[circle, red, fill, inner sep=1.5pt, label=below right:$\textcolor{red}{u_6}$,label=center:$\color{red}{\odot}$](b){};
\draw (13,0) node[circle, red, fill, inner sep=1.5pt, label=below right:$\textcolor{red}{u_9}$, label=center:$\color{red}{\odot}$](b){};
\draw (14,0) node[circle, red, fill, inner sep=1.5pt, label=below right:$\textcolor{red}{u_{12}}$, label=center:$\color{red}{\odot}$](b){};
\draw (11,1) node[circle, fill, inner sep=1.5pt, label=below right:$u_2$,label=center:$\odot$](b){};
\draw (12,1) node[circle, fill, inner sep=1.5pt, label=below right:$u_5$,label=center:$\odot$](b){};
\draw (13,1) node[circle, fill, inner sep=1.5pt, label=below right:$u_8$,label=center:$\odot$](b){};
\draw (14,1) node[circle, red, fill, inner sep=1.5pt, label=below right:$\textcolor{red}{u_{11}}$, label=center:$\color{red}{\odot}$](b){};
\draw (11,2) node[circle, fill, inner sep=1.5pt, label=below right:$u_1$, label=center:$\odot$](b){};
\draw (12,2) node[circle, fill, inner sep=1.5pt, label=below right:$u_4$, label=center:$\odot$](b){};
\draw (13,2) node[circle, fill, inner sep=1.5pt, label=below right:$u_7$,label=center:$\odot$](b){};
\draw (14,2) node[circle, red, fill, inner sep=1.5pt, label=below right:$\textcolor{red}{u_{10}}$, label=center:$\color{red}{\odot}$](b){};

\end{tikzpicture}
\end{center}
\end{minipage}
\caption{Additional grid points $\bullet$ are added to collocate the BVP. The least-square approach leads to a rectangular system matrix with more rows than columns.}
\label{fig:stsolls}
\end{figure}

The middle figure of Figure \ref{fig:stsolls} shows how the additional grid points ($\bullet$) are laid out. Note that we are only interested in the solutions located at the main grid points ($\odot$). From the figure, we add 17 grid points. The 12 black dots at the interior of the domain are the PDE locations, and five red dots are for boundary conditions. In the least-square version, the system matrix of \eqref{eq:spmatdistmodified} will be $29 \times 12$ in size and the right hand side is of size $29 \times 1$. Out of 17 additional rows, entries of the 12 additional PDE rows can be obtained by interpolating the PDE from the main grid. Likewise, entries of five additional boundary rows can be computed by interpolating boundary values from the main grid at the south and east edges. Lastly, additional entries of the right-hand side are adjusted accordingly based on the values of $f(x,y)$ or $g(x,y)$ at those extra grid points.


\section{Corichi and Singh Model of the Schwarzschild Interior}\label{CS-M}

As was done in [\refcite{Yonika_2019}], attention is paid towards the loop-quantized description of the Schwarzschild interior presented in~[\refcite{Corichi_2016}] for a Schwarzschild black hole of mass $m$. Loop quantization of the Hamiltonian constraints uses the connection coefficients $b$ and $c$, defining  
\begin{equation}\label{dbdc}
\delta_b = \frac{\sqrt{\Delta}}{2 m}, ~~~ \mathrm{and} ~~~~ \delta_c = \frac{\sqrt{\Delta}}{{L_o}} ~~~ 
\end{equation}
where $\Delta$ denotes the minimum area eigenvalue in loop quantum gravity, i.e. $\Delta = 4 \sqrt{3} \pi \gamma \lp^2$ with $\lp$ being the Planck length. The CS model Hamiltonian constraint is presented as a discretized equation over the two triad variables $\mu$ and $\tau$. With the wave function of the black hole $\Psi$, defined over discrete values in these two triad variables. Quantitatively, this is represented as:
\begin{eqnarray}\label{CS}
&& \nonumber (\sqrt{|\tau|} + \sqrt{|\tau + 2 \dc|}) \left(\Psi_{\mu + 2 \db, \tau + 2 \dc} - 
\Psi_{\mu - 2 \db, \tau + 2 \dc} \right) \\
&& \nonumber \hskip-0.2cm + \tfrac{1}{2\db} \, (\sqrt{|\tau + \dc|} - \sqrt{|\tau -  \dc|}) \bigg[(\mu+2\db)\Psi_{\mu + 4 \db, \tau} \\
&& \nonumber \hskip0.4cm ~~~~~ +  (\mu-2\db) 
\Psi_{\mu - 4 \db, \tau}  - 2 \mu (1 + 2 \gamma^2 \db^2)  \Psi_{\mu,\tau} \bigg]\\
&& \nonumber \hskip-0.2cm + (\sqrt{|\tau|} + \sqrt{|\tau - 2 \dc|}) \left(\Psi_{\mu - 2 \db, \tau - 2 \dc} - 
\Psi_{\mu + 2 \db, \tau - 2 \dc} \right) \\
&& = 0 ~,
\end{eqnarray}
with $\gamma$ being the Immirzi parameter. 

As it was shown in~[\refcite{Yonika_2018}], Eqn.~\ref{CS} has two separate stability constraints that are recovered through von-Neumann stability analysis. The first $\gamma\db \rightarrow 0$ is easily implementable, and the second $\mu < 4\tau$ requires careful selection of the placement of the initial conditions. Setting $\gamma\db=0$, and making the transformation that $2\dc=2\db=1$ transforms Eqn.~\ref{CS} into a more suitable form for implementation into a numerical codebase for computation. This results in, 
\begin{eqnarray}\label{CS-mod}
&& \nonumber (\sqrt{|\tau|} + \sqrt{|\tau + 1|}) \left(\Psi_{\mu + 1, \tau + 1} - 
\Psi_{\mu - 1, \tau + 1} \right) \\
&& \nonumber \hskip-0.2cm + \, (\sqrt{|\tau + \tfrac{1}{2}|} - \sqrt{|\tau -  \tfrac{1}{2}|}) \bigg[(\mu+1)\Psi_{\mu + 2, \tau} \\
&& \nonumber \hskip0.4cm ~~~~~ +  (\mu-1) 
\Psi_{\mu - 2, \tau}  - 2 \mu \Psi_{\mu,\tau} \bigg]\\
&& \nonumber \hskip-0.2cm + (\sqrt{|\tau|} + \sqrt{|\tau - 1|}) \left(\Psi_{\mu - 1, \tau - 1} - 
\Psi_{\mu + 1, \tau - 1} \right) \\
&& = 0 ~.
\end{eqnarray}
Throughout the literature~[\refcite{Yonika_2018,Yonika_2019}], computations with the CS model after implementation of the discovered stability constraints results in an important feature of symmetry in both $\mu$ and $\tau$. The solution demonstrates the quantum gravity prediction of a {\em big bounce} event, signalling a black hole to white hole transition~[\refcite{Haggard_2015,Bianchi_2018,Olmedo_2017}] and the assertion of singularity resolution presented by~[\refcite{Corichi_2016}].

\section{Space-Time Collocation Framework for CS Model}\label{ST-CS}

Furthering the work done in~[\refcite{Yonika_2019}] of the collocation-inspired framework, a {\em space-time} approach over both of the present triad variables is desired. Wherein the generalized collocation framework can be thought of as representing the problem as a linear system of equations, i.e. 
\begin{equation}
    \mathbf{\Phi}_x \vec{\omega} = \vec{f}~.
\end{equation}
Where the motivation is to solve the linear system for a collection of weights $\omega_i$, such that an accurate answer can be reconstructed in a domain by, 
\begin{equation}
    \sum_i \Phi_i(x)\omega_i = f(x)~.
\end{equation}
This can be easily extended into a STCM implementation by taking the tensor product over a separate spatial and temporal part. That is, considering a spatial basis $\Phi(x)$ and a temporal basis $\Theta(t)$, a STCM linear system can be represented by,
\begin{equation}
    (\mathbf{\Phi}_x\otimes\mathbf{\Theta}_t)(\vec{\omega}_x\otimes\vec{\omega}_t) = g(t,x)
\end{equation}
Applying this to Eqn.~\ref{CS-mod} implies a tensor product for each discrete value of $\Psi$. This results in the collocation matrix elements being given by: 
\begin{eqnarray}\label{CS_vander}
&& \nonumber (\sqrt{|\tau_n|} + \sqrt{|(\tau_n+1)|}) \left(\theta_l (\tau_n+1)\phi_m (\mu_k+1) - 
\theta_l (\tau_n+1)\phi_m (\mu_k-1) \right) \\
&& \nonumber \hskip-0.2cm + \, (\sqrt{|(\tau_n+\tfrac{1}{2})|} - \sqrt{|\tau_n-\tfrac{1}{2})|}) \bigg[(\mu_k+1)\theta_l(\tau_n)\phi_m(\mu_k+2) \\
&& \nonumber \hskip0.4cm ~~~~~ +  ( \mu_k-1) 
\theta_l(\tau_n)\phi_m(\mu_k-2)  - 2 \mu_k \theta_l(\tau_n)\phi_m(\mu_k) \bigg]\\
&& \nonumber \hskip-0.2cm + (\sqrt{|\tau_n|} + \sqrt{|(\tau_n-1)|}) \left(\theta_l(\tau_n-1)\phi_m(\mu_k-1) - 
\theta_l(\tau_n-1)\phi_m(\mu_k+1) \right) \\
&& \equiv \Phi_{lm}(\mu_k,\tau_n).
\end{eqnarray}
Where each element refers to a spatial and temporal basis index, and also a spatial and temporal node. Initial conditions are included with the conventional process of row/column injection and take the form of two Gaussian peaks, symmetric about $\mu=0$.
\subsection{Choice of Basis}\label{ST-CS-B}

As shown in~[\refcite{Yonika_2019}] a basis in $\mu$ was used that converged to smooth behavior at large values of the triad variable. In that work, inspiration from spectral collocation methods was taken and a modified Fourier basis was used. The basis took the form of a Fourier basis with exponentially decaying ``wave number'' $k(\mu)$ and was inspired by work done previously in~[\refcite{Connors_2006}]. This remained unchanged in the space-time approach, with the spatial basis reliably taking the form:
\begin{equation}\label{mu-basis}
    \phi_m(\mu_k) = {\rm exp}(ime^{-|\tfrac{\mu_k}{M}|{\rm exp}(\tfrac{2m}{M})})
\end{equation}
As previously, the symmetry of the solution allows us to treat the basis function as an even interpolation where the basis is represented by indices $m\in[0,M]$. And a similar distribution of $\mu$ nodes are selected for the spatial sequence, i.e. following the distribution for $K$ nodes:
\begin{equation}\label{mu-nodes}
    \mu_k = \mu_{k-1} + \lfloor 1 + \left(\tfrac{2\mu_{k-1}}{K}\right)^2\rfloor
\end{equation}
For the temporal basis, promising results were generated on a similar modified Fourier basis and also a polynomial basis. Results will be discussed with each basis choice of the following,
\begin{align}\label{tau-basis}
    \theta_l(\tau_n) =& |\frac{\tau_n}{{\rm T}}|^{\tfrac{l}{L}} & \theta_l(\tau_n) &= {\rm exp}(ile^{-|\tfrac{\tau_n}{{\rm T}}|})
\end{align}
Taking ${\rm T}$ as the largest $\tau$ value in our domain and including basis elements referenced by positive indices, i.e. $l\in[0,M]$. The nodal distribution in $\tau$ for $N$ nodes was taken to be a uniform distribution on the grid, 
\begin{equation}\label{tau-nodes}
    \tau_n = {\rm T}\frac{n+1}{N}~.
\end{equation}
An example set of nodes generated with Eqn's.~\ref{mu-nodes},\ref{tau-nodes} is demonstrated below in Fig.~\ref{fig:nodes}, only a subset of elements is shown for clarity.
\begin{figure}[htb!]
    \centering
    \includegraphics[width=.5\linewidth]{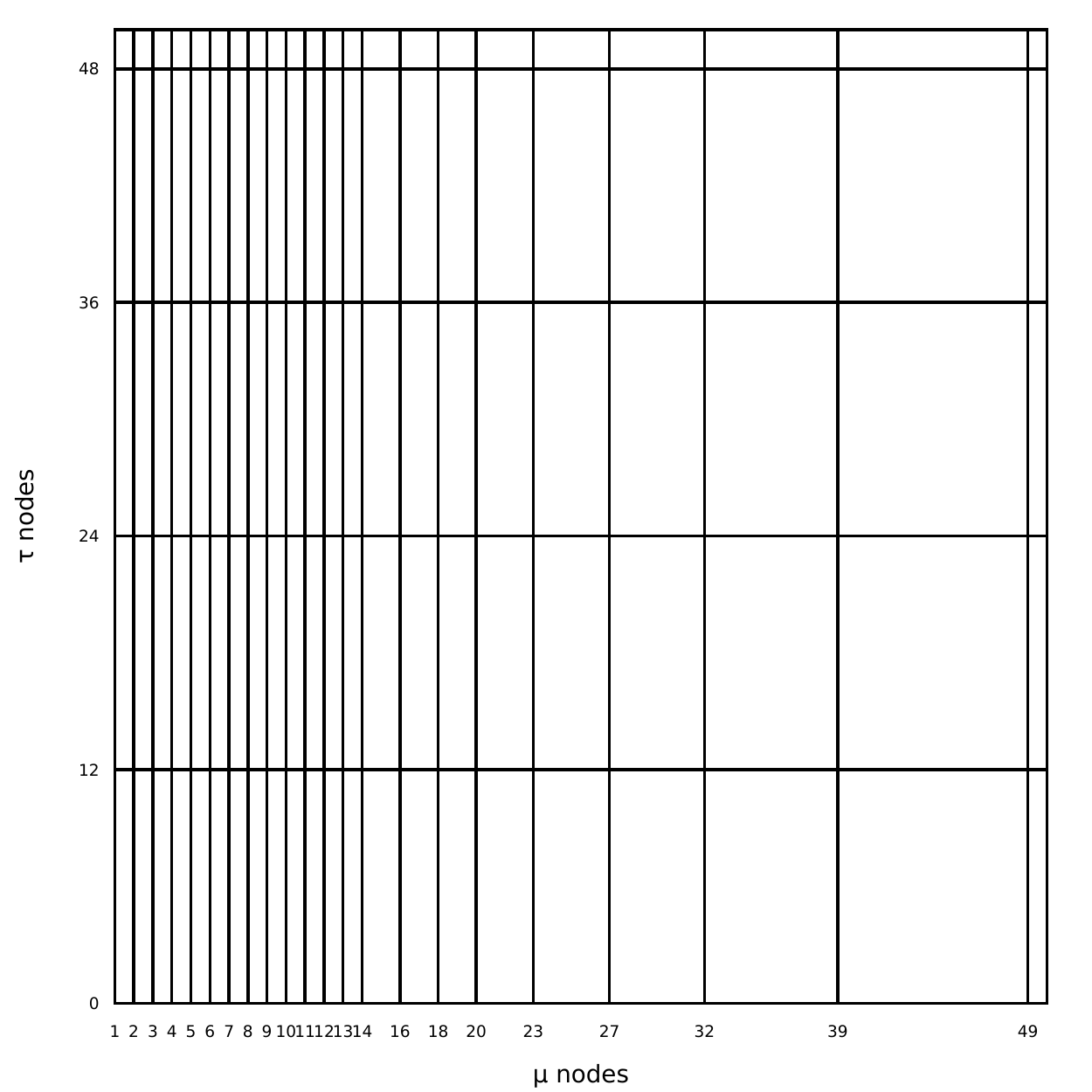}
    \caption{Example collection of $\mu$ and $\tau$ nodes defined sparsely upon the positive quadrant of a square grid}
    \label{fig:nodes}
\end{figure}\newpage
\subsection{Least-Squares Approach}\label{ST-CS-LS}

Performing the STCM on a uniquely-determined system, $L=N=M=K$ led to divergent results due to ill-conditioning of the linear system. Resulting in a loss of singularity resolution, and unstable repetitions of the imposed initial conditions through the domain. It was suspected that the nature of the chosen non-orthogonal bases led to $\Phi_{lm}(\mu_k,\tau_n)$ becoming rank-deficient; the details of such are included in later discussion in the context of this current work. A {\em least-squares} approach, as discussed in~[\refcite{houstis1978collocation}], to the linear system problem was implemented as a way to enforce stability-preserving properties. The collocation matrix $\Phi_{lm}(\mu_k,\tau_n)$ was changed to represent an overdetermined system, i.e. $K,N > M=L$. A vector solution for the weights was then picked by minimizing the residual, 
\begin{equation}\label{least-squares}
    \underset{\vec \omega}{{\rm min}}\left(||\mathbf \Phi \vec \omega - \vec g||\right)
\end{equation}

Additionally, to disrupt ill-conditioning under increasing basis size $M$; scaling of the collocation matrix was performed by dividing a common factor in the rows responsible for the initial and boundary conditions. The common factor was taken to be the maximum value of the initial conditions.

\section{Results}\label{Results}

The least-squares approach led to the production of expected results for the CS model under an evolution in a square domain.
\begin{figure}[htb!]
    \centering
    \includegraphics[width=.45\linewidth]{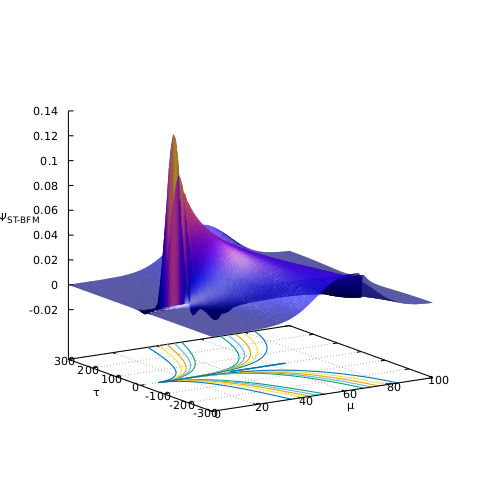}
    \includegraphics[width=.45\linewidth]{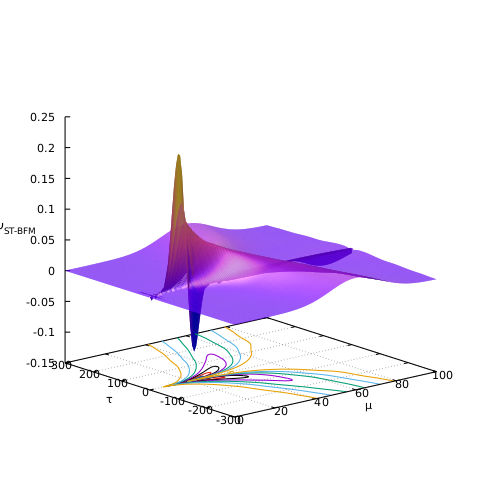}
    \caption{Surface plot of STCM generated solutions with polynomial basis in $\tau$ (left), and modified-Fourier basis in $\tau$ (right).}
    \label{fig:surface}
\end{figure}\\

The expected features are observable in the surface plots presented in Fig.~\ref{fig:surface}, namely a trajectory for the Gaussian wave packet that approaches $\mu = 0$ as $\tau \rightarrow 0$. Concurrently, the packet gets narrower throughout the evolution and exhibits rapidly fluctuating behavior at the location of the classical singularity at $\tau = 0$. The quantum gravity predictions of avoiding the classical collapse of the wave packet is observed, and bounce behavior is noted by the symmetric nature about $\tau = 0$.
\subsection{Exact Solution Comparisons}\label{R-E}

Error can be computed via comparison with the exact solution, represented by solving Eqn.~\ref{CS} in an iterative-stepping fashion. Computing the residual across slices of constant $\tau$ between the exact PDE solution and the reconstructed solution achieved by the STCM approach, allows a quantitative measurement of the error to be calculated. Presented are the traditional metrics of $L_2$ and $L_\infty$ norms of the residual, i.e.
\begin{align}\label{error-norms}
    L_2(\tau) =& \frac{\sum_i (\Psi_{STCM}(\mu_i,\tau)-\Psi_{\mu_i,\tau})^2}{\sum_i \Psi_{\mu_i,\tau}^2}  &  L_\infty(\tau) &= \underset{\mu}{{\rm max}}(\Psi_{STCM}(\mu_i,\tau)-\Psi_{\mu_i,\tau})
\end{align}
\begin{figure}[htb!]
    \centering
    \includegraphics[width=.45\linewidth]{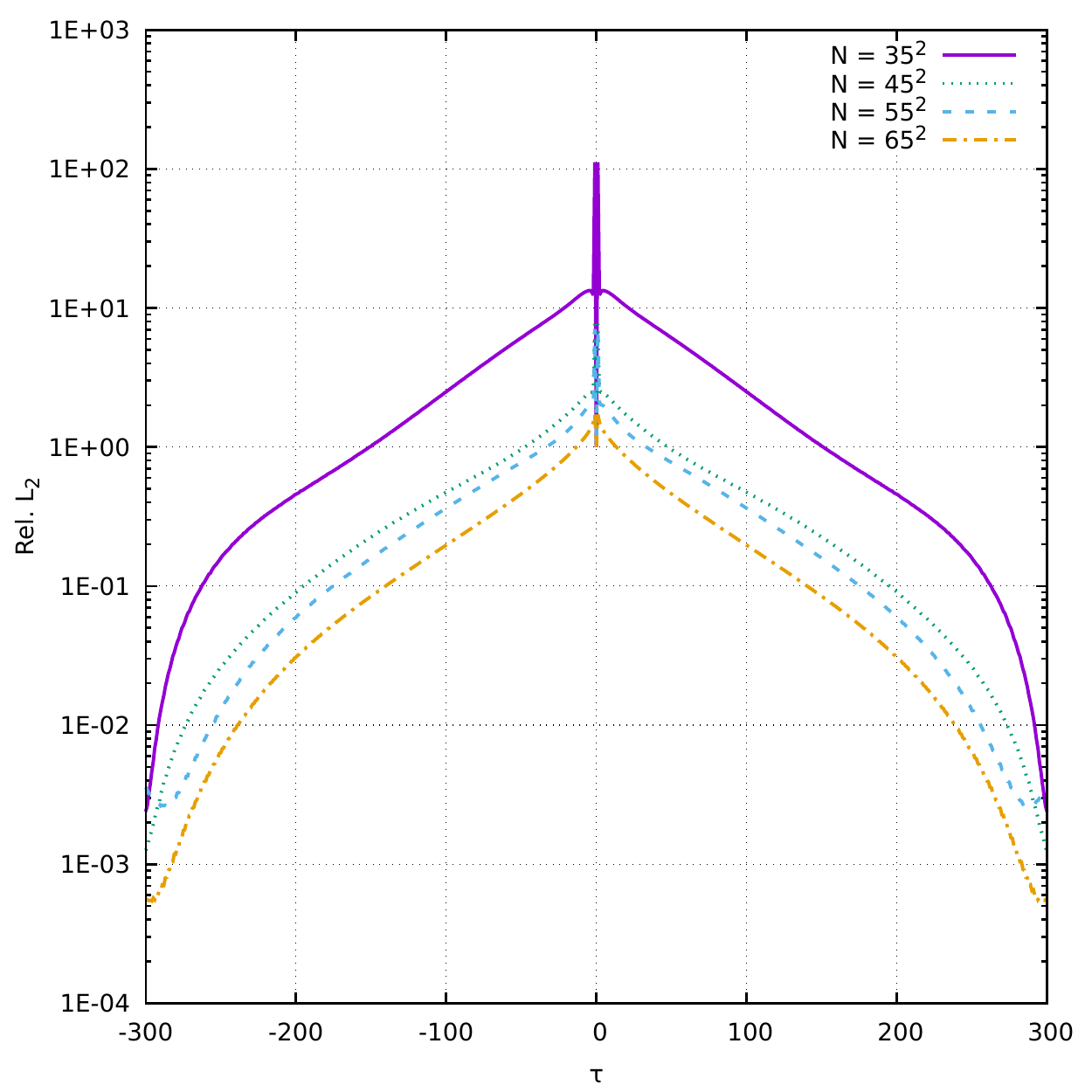}
    \includegraphics[width=.45\linewidth]{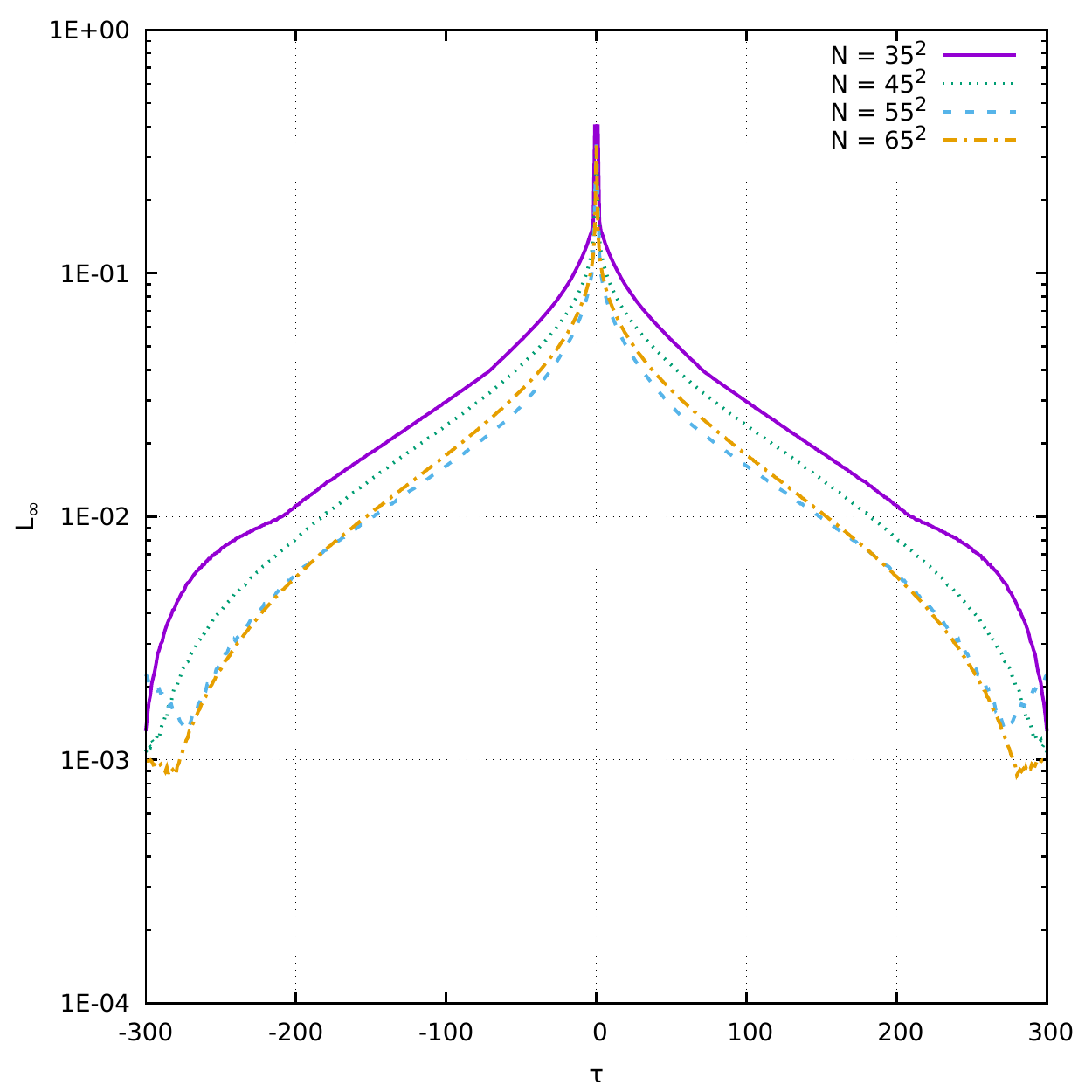}
    \caption{Relative $L_2$-norm error for varying basis size $N$ (left) along with $L_\infty$-norm error (right) for the polynomial basis in $\tau$}
    \label{fig:error-poly}
\end{figure}\\

\begin{figure}[htb!]
    \centering
    \includegraphics[width=.45\linewidth]{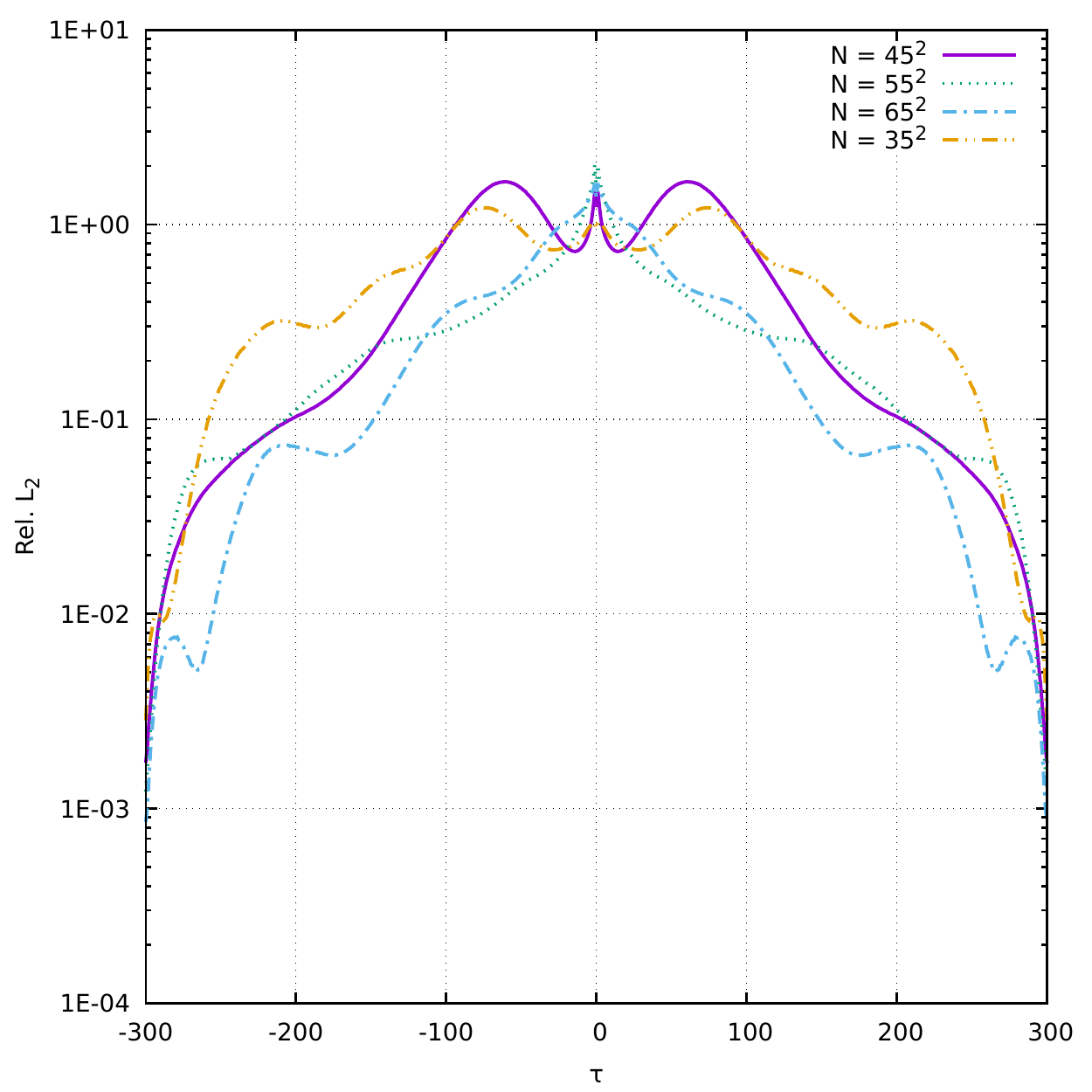}
    \includegraphics[width=.45\linewidth]{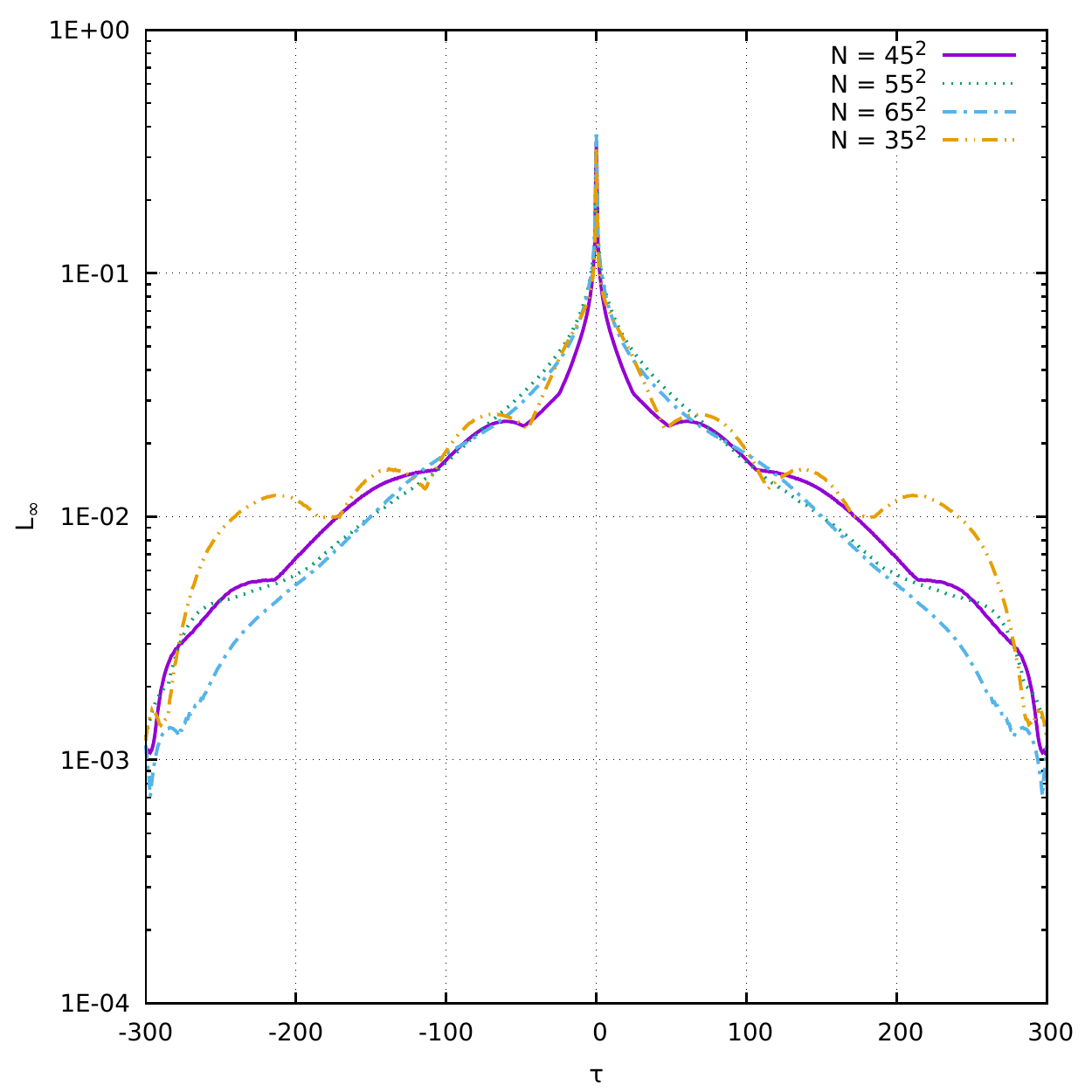}
    \caption{Relative $L_2$-norm error for varying basis size $N$ (left) along with $L_\infty$-norm error (right) for the modified-Fourier basis in $\tau$}
    \label{fig:error-four}
\end{figure}

Shown in Fig.~\ref{fig:error-poly} is an overall decrease in $L_2$ and $L_\infty$ error measures as the size of the basis grows, as expected. A rapid accumulation of error for both the $L_2$ and $L_\infty$ metrics is present around $\tau=0$; due to the behavior of the polynomial basis function vanishing to zero at that point. This feature can be reduced by the use of the modified-Fourier basis presented in Eqn.~\ref{tau-basis}. The metrics for performance can be similarly computed as shown in Fig.~\ref{fig:error-four}, reducing the error around $\tau = 0$.
\newpage

\subsection{Parallel Benefit}\label{R-P}

To demonstrate typical benefit to be expected through multi-threading; strong scaling tests were performed by keeping the problem size fixed, in terms of domain and basis size, and increasing the thread-count for the distributed sections of the simulation. 
\begin{figure}[htb!]
    \centering
    \includegraphics[width=.45\textwidth]{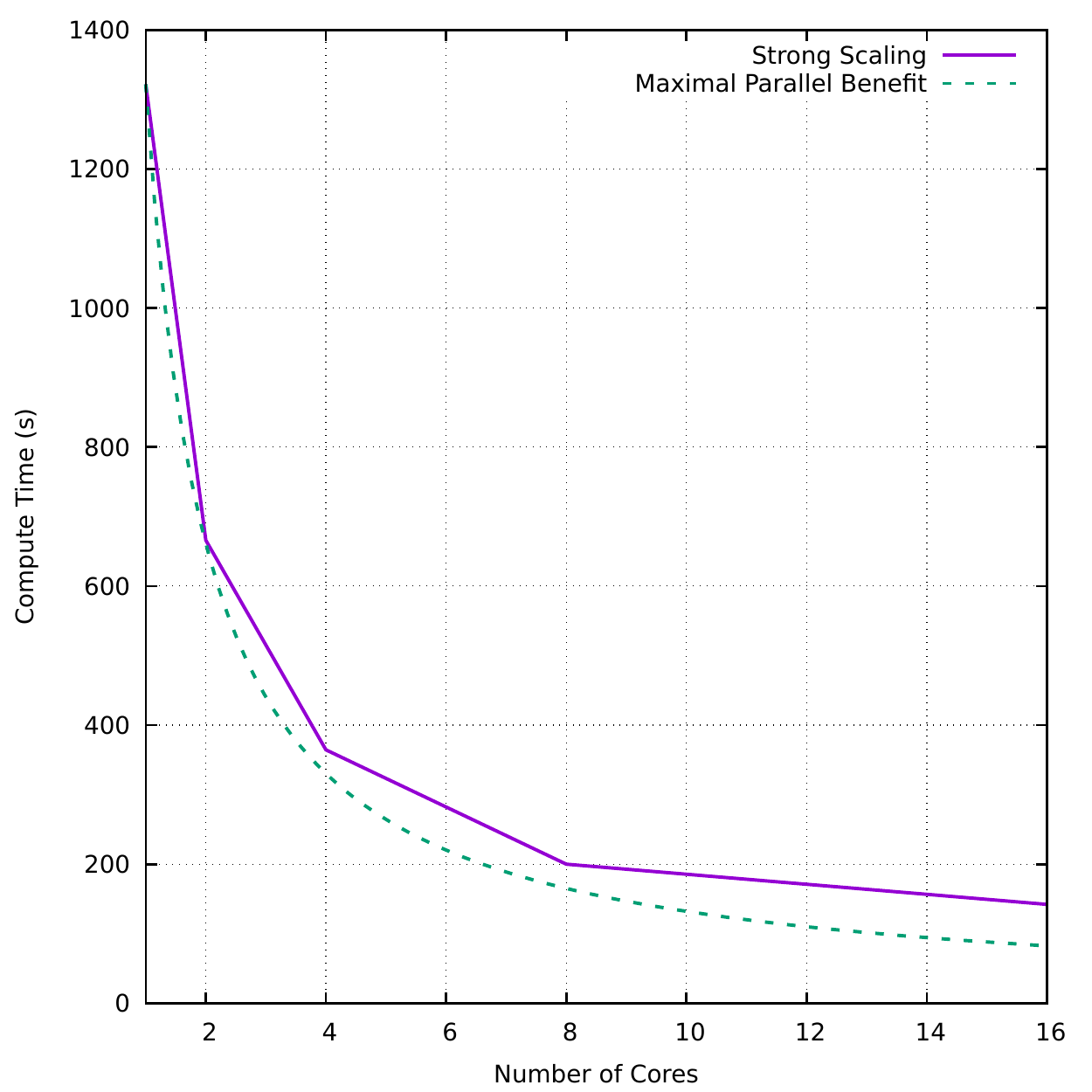}
    \caption{Strong scaling benchmarking of STCM computed solution}
    \label{fig:parallel}
\end{figure}\\

As can be seen in Fig.~\ref{fig:parallel}, parallel-computing provides demonstrable benefit due to the {\em simultaneous} nature of the underlying computations.
\subsection{Accuracy of Method}\label{R-A}

\begin{figure}[htb!]
    \centering
    \includegraphics[width=.45\linewidth]{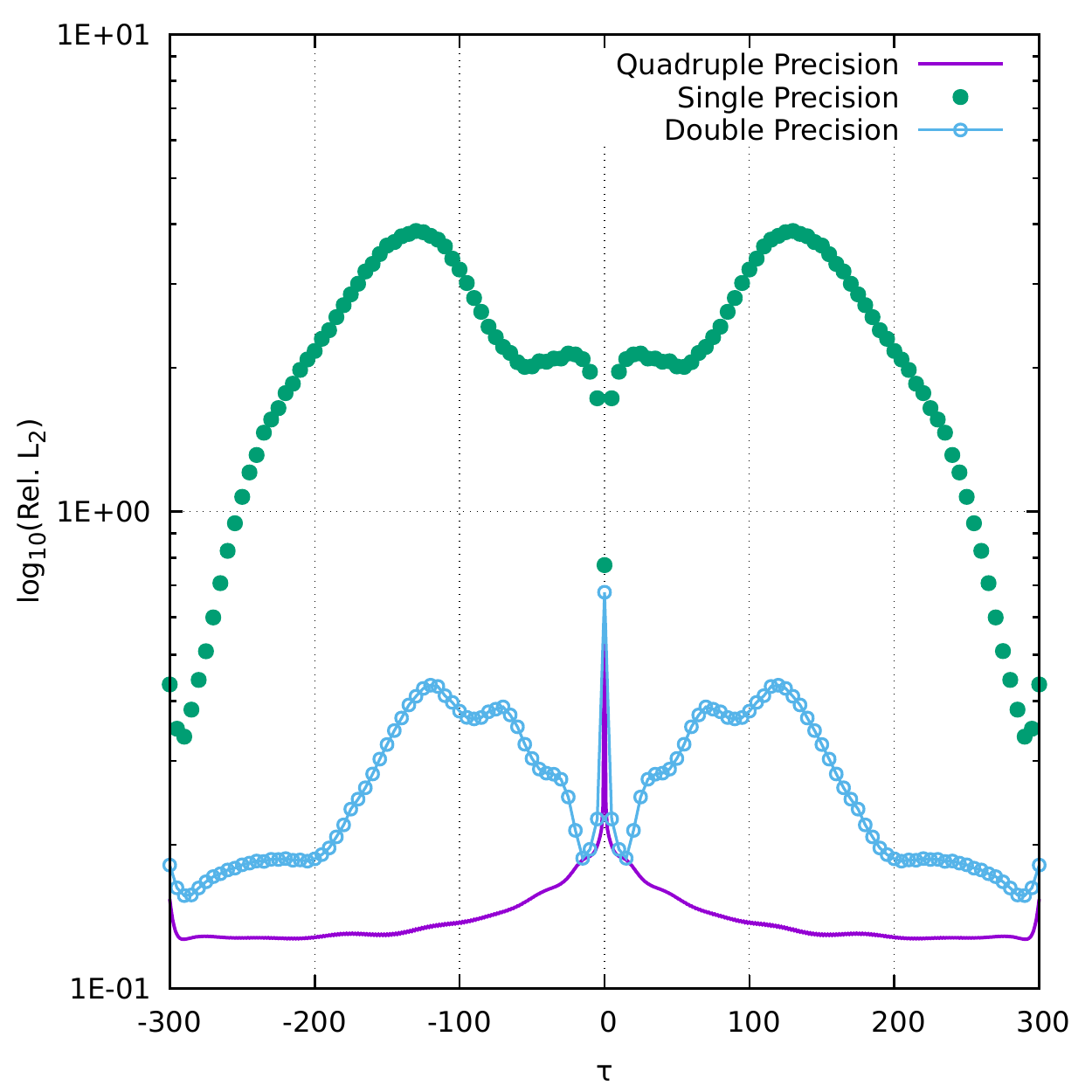}
    \includegraphics[width=.45\linewidth]{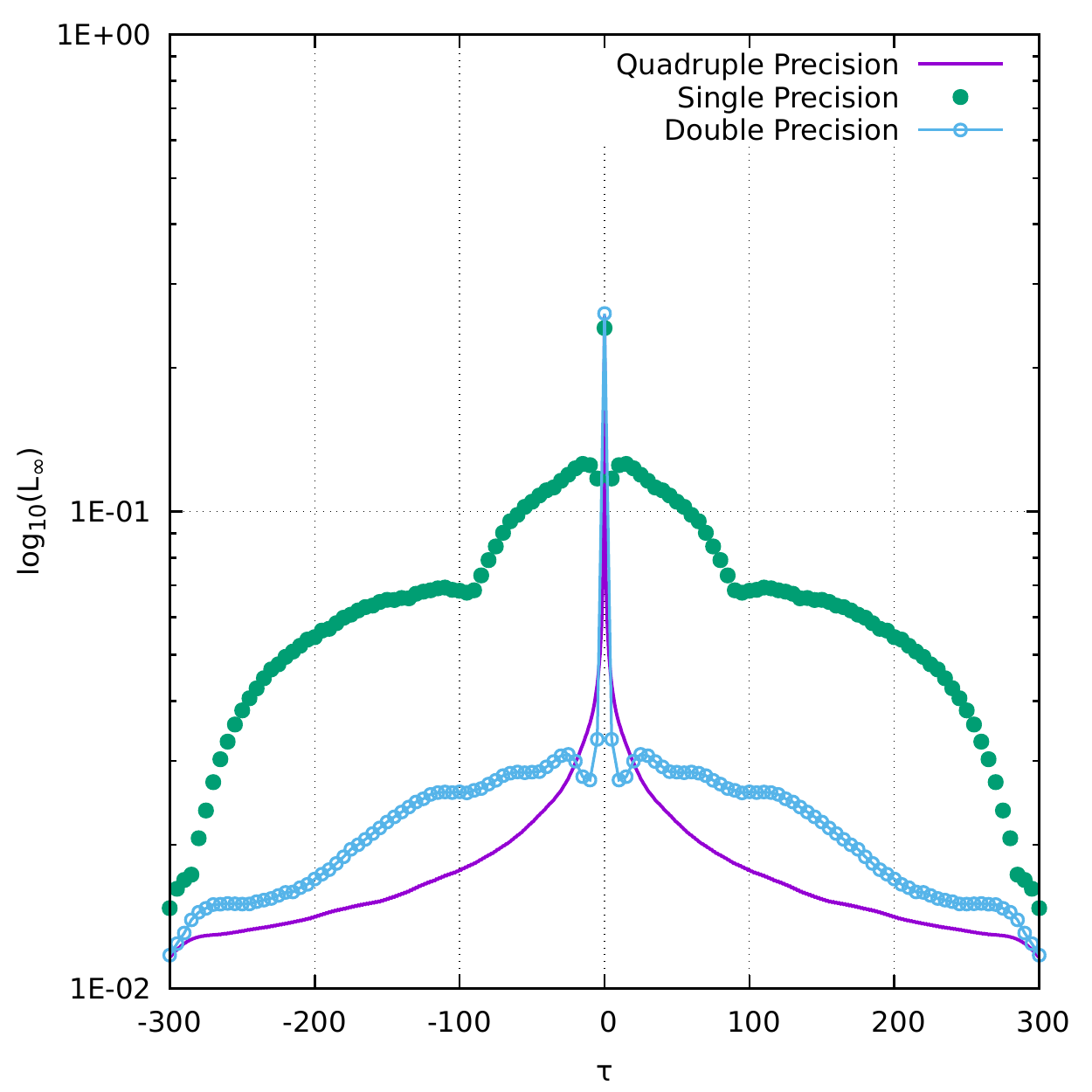}
    \caption{Relative $L_2$-norm error and $L_\infty$-norm error for various levels of finite precision}
    \label{fig:precision}
\end{figure}

Additional insight can be gathered into the accuracy of the space-time method by analyzing the behavior upon successive calculations of increasing levels of utilized finite-precision arithmetic. It is expected that as the level of precision is increased, e.g. from single to double precision, the error should decrease. This is due to added sensitivity towards minuscule changes in number representation by a reduction in magnitude of the respective machine-precision $\epsilon_{\rm{M}}$ as the Byte assignment of the computations is increased. Fig.~\ref{fig:precision} shows this phenomena, where each computed solution has been compared to the exact solution through the iterative-stepping method. Thus giving a measure of the round-off error involved in the space-time method.
\newpage

The results of the precision testing showed two interesting points. First, higher levels of precision do not necessarily aid in the computation of an accurate solution. This was seen when the number of basis elements grew too large, resulting in a loss of accuracy for higher-precision solutions. It is suspected that the basis converges more rapidly to an orthonormal basis with a smaller number of basis elements for higher-precision computations. Secondly, the space-time least-squares approach is more tolerant to lower levels of precision compared with the hybrid approach and iterative-stepping discussed in~[\refcite{Yonika_2019}]. With accurate solutions reliably computed in double precision, as opposed to quadruple.

\begin{figure}[htb!]
    \centering
    \includegraphics[width=.45\linewidth]{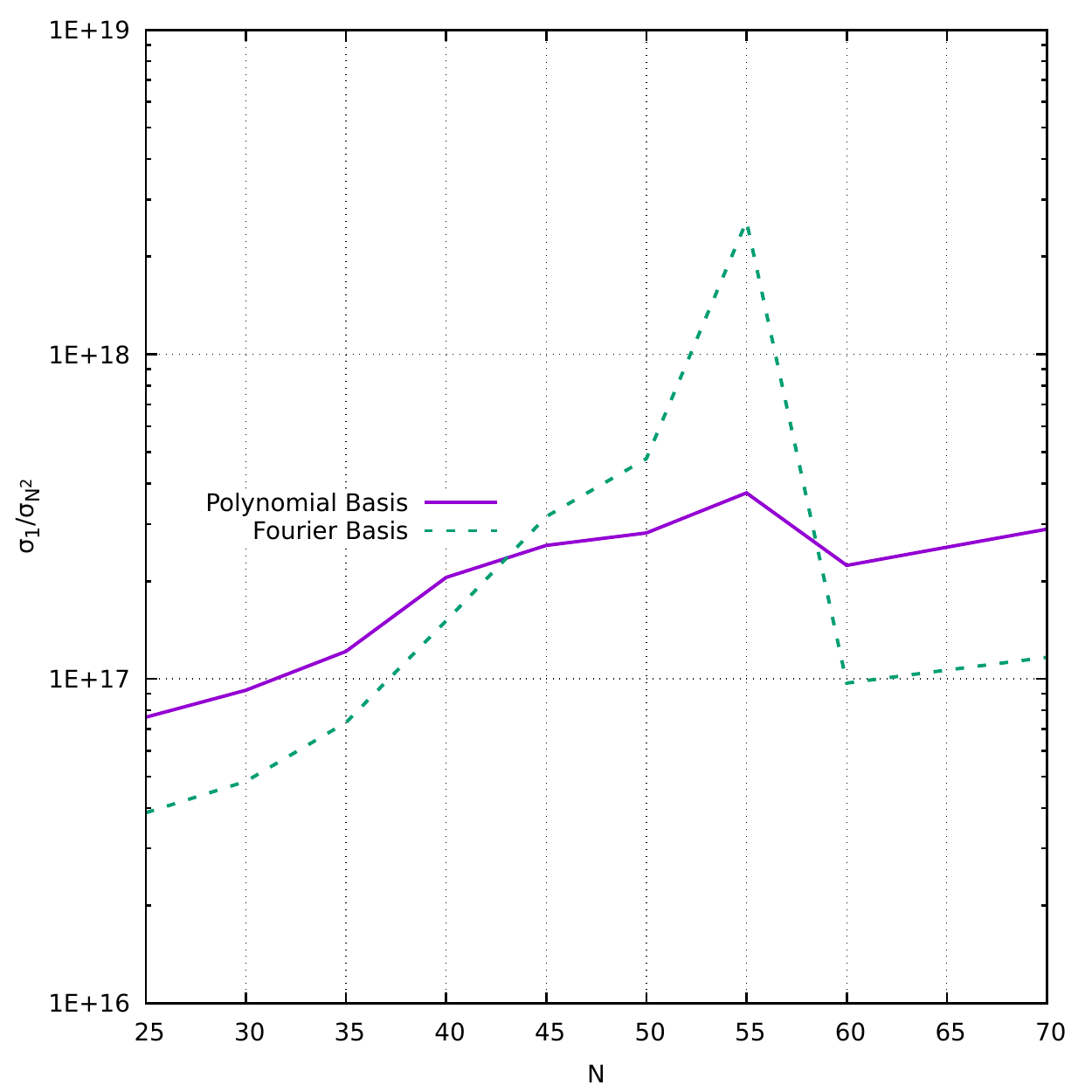}
    \includegraphics[width=.45\linewidth]{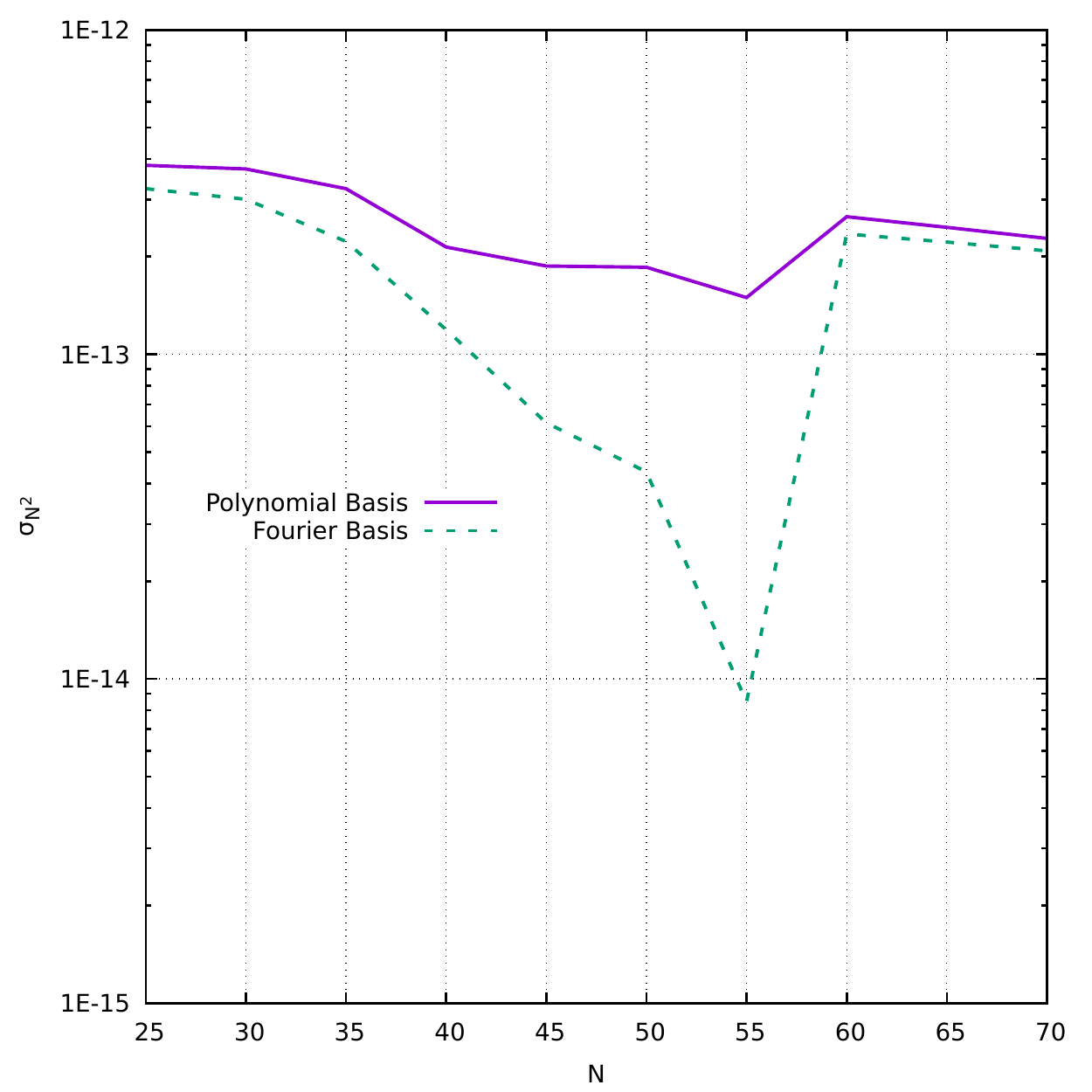}
    \caption{Condition number (left) for variable-sized basis smallest singular value (right) for variable sized basis of collocation matrix.}
    \label{fig:svd1}
\end{figure}

In order to provide additional insight into the accuracy of the space-time approach, SVD analysis can be performed to determine some relevant features. The condition number, computed as the ratio of largest to smallest singular value $\kappa_2 = \tfrac{\sigma_1}{\sigma_{N^2}}$, gives a measure of the order of accuracy involved in inverting the interpolation matrix $\mathbf\Phi$. Therefore $\kappa_2$ can serve as an estimate of the upper bound on the error of the space-time method. For dense matrix representations, as we have through Eqn.~\ref{CS_vander}, the error introduced by numerically solving the solution has an upper bound determined as $\epsilon_{\rm{S}} = ||\mathbf\Phi\vec\omega-\vec g||_2 \propto \mathcal{O}(\kappa_2\epsilon_{\rm{M}})$. Whereas the smallest singular value $\sigma_{N^2}$ denotes the order of the computed $N^2$ weight. Through this metric, it can be seen if the computations required to solve the linear system require sub-$\epsilon_{\rm{M}}$ arithmetic. These metrics are shown in Fig.~\ref{fig:svd1} for each choice of $\tau$ basis in Eqn.~\ref{tau-basis}.
\\

\begin{figure}[htb!]
    \centering
    \includegraphics[width=.45\linewidth]{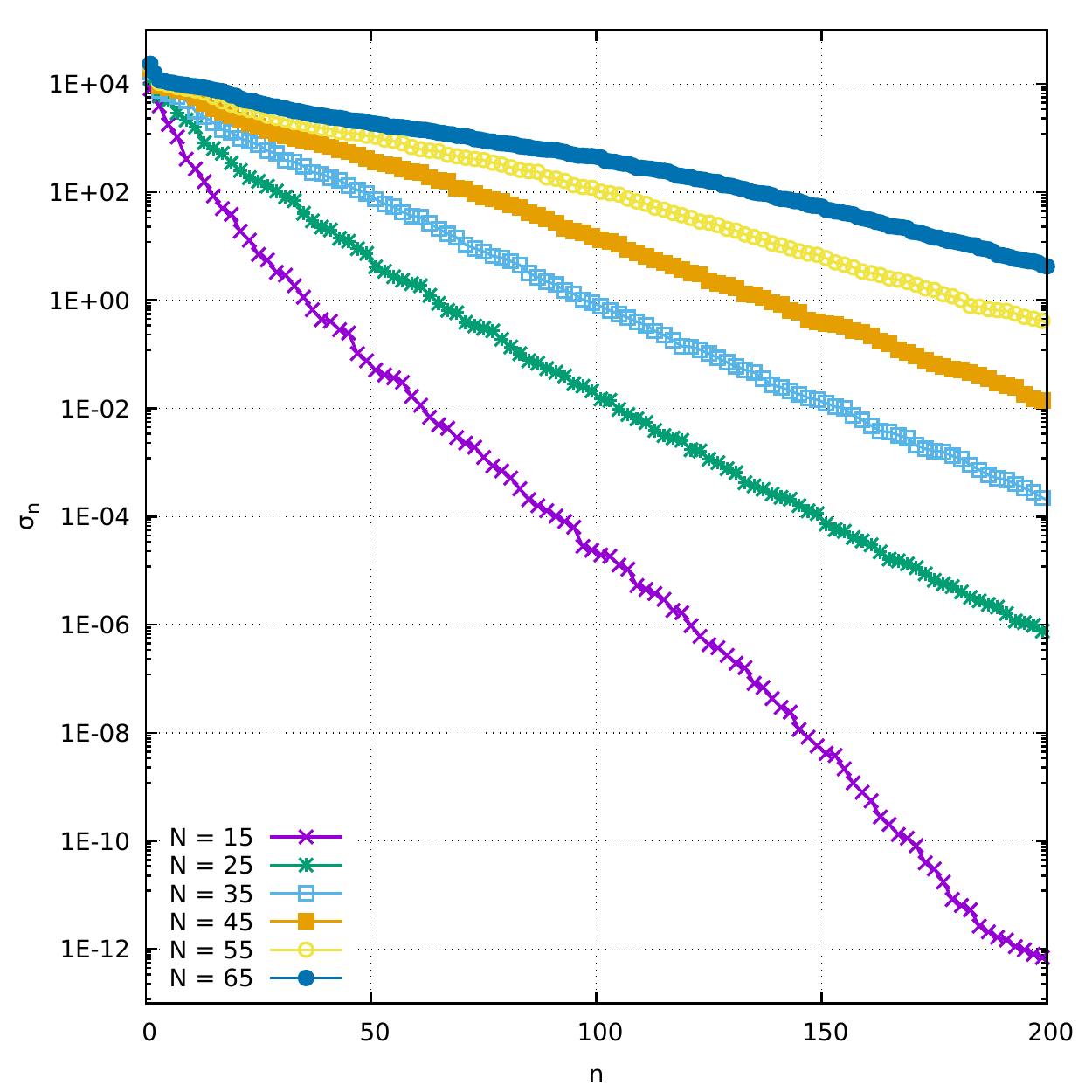}
    \includegraphics[width=.45\linewidth]{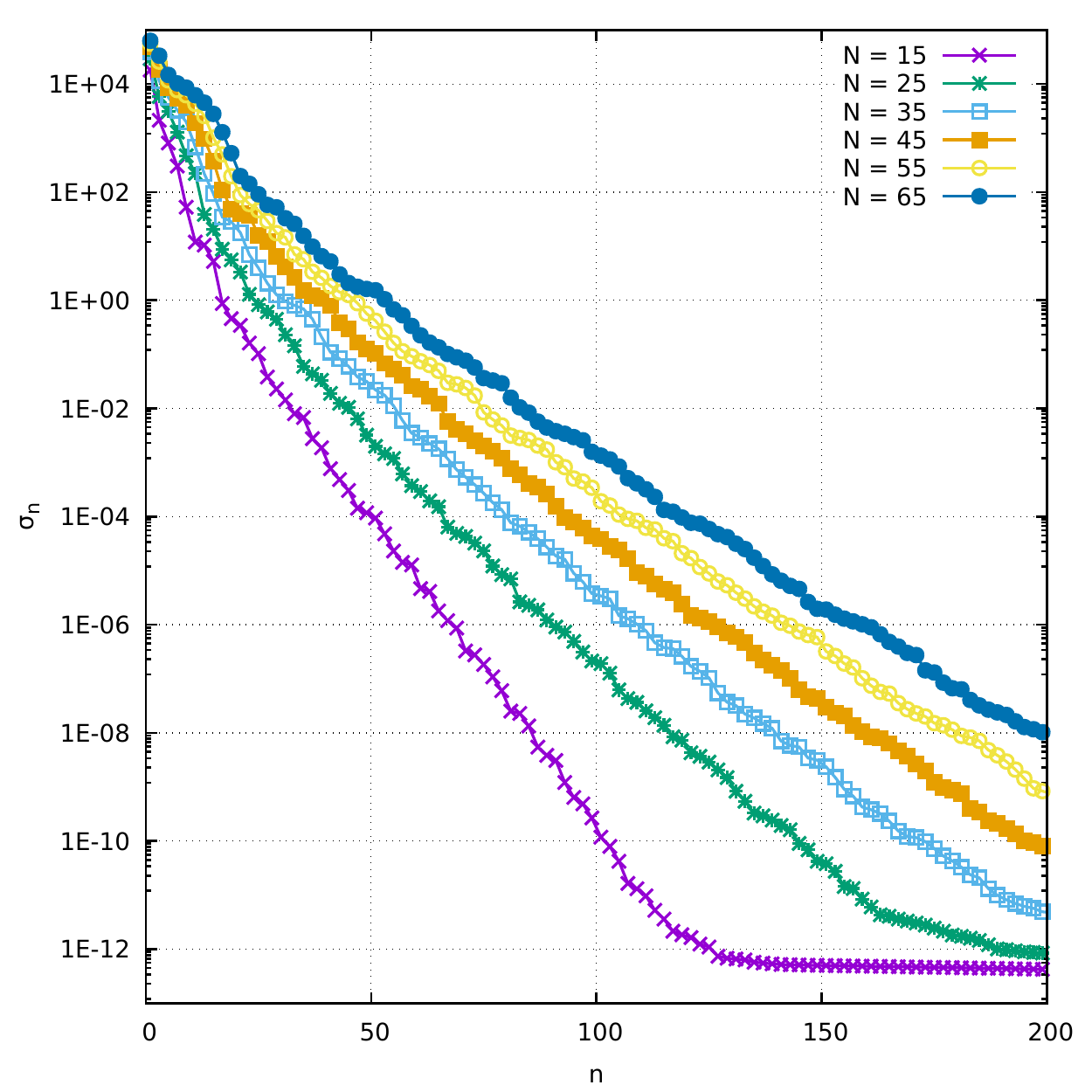}
    \caption{Spectra of singular values for modified-Fourier (left) and polynomial (right) basis in $\tau$}
    \label{fig:svd2}
\end{figure}

It is also useful to note the exponential decay of the singular values, a feature typical of approximation via exponential sums. Where the exponential decay of $\sigma$ corresponds to a feature where each additional basis element improves the accuracy of the solution; and larger rates of decay of $\sigma$ imply faster convergence of the finite sum of basis elements~[\refcite{BEYLKIN200517}]. This is shown for both $\tau$ basis choices from Eqn.~\ref{tau-basis} for varying basis size $N^2$ in Fig.~\ref{fig:svd2} with the polynomial basis demonstrating a more rapid decay of the singular values over the modified-Fourier basis of equivalent size.
\\

Note that each respective choice of $\tau$ basis examined in this paper provides different characteristics that affects the feasibility of finding a solution via the space-time approach. Through Fig.~\ref{fig:svd1}, it is observed that the modified-Fourier basis often imposes better conditioning on the linear system. However, as shown in Fig.~\ref{fig:svd2} the polynomial basis converges quicker as the size of the basis is increased.

Our numerical experiments show that the choice of non-orthogonal bases that captures the physical model of the problem does lead to an ill-conditioned non-symmetric square system. The system approximates the discrete PDE and boundary condition operators in the space-time domain. By using SVD, the decay of singular values shows the number of orthogonal bases in the form of rank-one matrices needed to approximate the system operator $\Phi$ accurately. As an example, for the polynomial in $\tau$ case with $N=25$ (i.e., $625 \times 625$ square system) shown in Fig.~\ref{fig:svd2}, it takes about $n=150$ bases to approximate the system operator $\Phi$ to $\mathcal{O}(10^{-12})$ accuracy. Thus, the operator $\Phi$ can be approximated by a $625 \times 150$ rectangular matrix (its lower-rank approximation) instead of the full size $625 \times 625$. This essentially discards columns that do not matter anymore since the singular values already flatten out.

Due to its computational costs as the degree of freedom becomes larger, we want to avoid performing SVD to generate the orthogonal bases. Instead, we are utilizing a least-squares technique to generate a rectangular system (more rows than columns). One may consult~[\refcite{Ake_Bjorck_book}] for an extensive study in this field. The quadruple precision can be used to delay any other conditioning issue that may arise, which cannot be alleviated by using the least-squares technique.

Finding orthogonal bases analytically with the Gram-Schmidt technique and a preconditioner of solving the square system will be left for future works. 

\section{Conclusion}
In this paper we developed and presented a space-time collocation-inspired approach towards solving a problem in quantum gravity regarding evolution equations in a loop-quantized Schwarzschild interior spacetime. The exact solution of the evolution equation took the form of a discrete {\em partial-difference equation}. The space-time method took heavy influence from the family of collocation methods by representing an assumed solution as a sum over a tensor product of independent basis functions. A linear least-squares approach was implemented in order to impose stricter adherence to stable solutions. Results were demonstrated with two different choices of basis in the time variable. Demonstrating a freedom of choice, and opportunity for an optimally chosen functional form of the basis inherent in collocation methods. When comparing the solution to the previously-used approach of iteratively stepping through a discretized grid and performing a set of recursive computations, issues that commonly arise and prohibit large-scale solutions to be feasibly computed were remedied. The collocation-inspired method provided a computationally-efficient yet accurate means to compute a solution upon a square grid; and due to the simultaneous nature of the approach, is readily scalable and provides an avenue for demonstrable benefit with standard high-performance computing techniques.
\\

{\em Acknowledgements}:  The authors would like to thank students participating in the University of Massachusetts Dartmouth Mathematical and Computational Consulting 2019 course and Dr. Scott Field for  fruitful discussion. G. K. acknowledges research support from NSF Grants PHY-1701284, DMS-1912716, and  ONR/ DURIP Grant No. N00014181255. 


\bibliographystyle{ws-ijmpc}
\bibliography{spacetime}

\end{document}